\documentclass[journal=jctcce,manuscript=article,layout=twocolumn]{achemso}
\setkeys{acs}{articletitle=true}

\usepackage[T1]{fontenc}
\usepackage{amsmath, amssymb, mathtools}
\usepackage{siunitx}
\usepackage{xcolor}
\usepackage{braket}
\usepackage[version = 3]{mhchem}
\usepackage{paralist}
\usepackage[para,online,flushleft]{threeparttable}
\usepackage[normalem]{ulem}
\usepackage{cuted}
\usepackage{rotating}

\newcommand{\onlinecite}[1]{\hspace{-1 ex}
\nocite{#1}\citenum{#1}}

\let\oldmaketitle\maketitle
\let\maketitle\relax

\title{Symmetry-adapted perturbation theory based on multiconfigurational wave function description of monomers}
\author{Micha{\l} Hapka}
\email{michal.hapka@uw.edu.pl}
\affiliation{Institute of Physics, Lodz University of Technology, ul.\ Wolczanska 219, 90-924 Lodz, Poland}
\alsoaffiliation{Faculty of Chemistry, University of Warsaw, ul.\ L.\ Pasteura 1, 02-093 Warsaw, Poland}
\author{Micha{\l} Przybytek}
\affiliation{Faculty of Chemistry, University of Warsaw, ul.\ L.\ Pasteura 1, 02-093 Warsaw, Poland}
\author{Katarzyna Pernal}
\affiliation{Institute of Physics, Lodz University of Technology, ul.\ Wolczanska 219, 90-924 Lodz, Poland}

\begin{document}

\twocolumn[
\begin{@twocolumnfalse}
\oldmaketitle
\begin{abstract}
We present a formulation of the multiconfigurational (MC) wave function symmetry-adapted perturbation theory (SAPT). The method is applicable to noncovalent interactions between monomers which require a multiconfigurational description, in particular when the interacting system is strongly correlated or in an electronically excited state. SAPT(MC) is based on one- and two-particle reduced density matrices of the monomers and assumes the single-exchange approximation for the exchange energy contributions. Second-order terms are expressed through response properties from extended random phase approximation (ERPA) equations. SAPT(MC) is applied either with generalized valence bond perfect pairing (GVB) or with complete active space self consistent field (CASSCF) treatment of the monomers. We discuss two model multireference systems: the H$_2\cdots$H$_2$ dimer in out-of-equilibrium geometries and interaction between the argon atom and excited state of ethylene. In both cases SAPT(MC) closely reproduces benchmark results.  Using the \ce{C2H4^*}$\cdots$Ar complex as an example, we examine second-order terms arising from negative transitions in the linear response function of an excited monomer. We demonstrate that the negative-transition terms must be accounted for to ensure qualitative prediction of induction and dispersion energies and develop a procedure allowing for their computation. Factors limiting the accuracy of SAPT(MC) are discussed in comparison with other second-order SAPT schemes on a data set of small single-reference dimers.
\end{abstract}
\end{@twocolumnfalse}
]


\section{Introduction}

Quantum chemistry offers two complementary approaches to study noncovalent interactions---the supermolecular approach and energy decomposition methods. The former is conceptually simple and capable of providing most accurate potential energy surfaces, e.g., for interpretation of experiments carried out in the cold- and ultracold regimes.~\cite{Bala:16,Dulieu:18,Pawlak:21} Decomposition methods allow insight into the nature of the interaction by partitioning the interaction energy into well-defined contributions. Both approaches have been extensively developed and benchmarked for systems which can be adequately described with single determinants.~\cite{Rezac:16,Patkowski:17} 
With the supermolecular method it is now possible to reach even submicro- or microhartree accuracy for small systems.~\cite{Patkowski:08,Jankowski:12,Bakr:13,Thibault:17,Przybytek:17} At the same time performing realistic energy decomposition for enzymes exceeding 3,000 atoms is also within reach.~\cite{Parrish:18} Energy decomposition schemes are not only interpretative tools, but can also provide potentials for quantitative predictions. The symmetry adapted perturbation theory (SAPT)\cite{Szalewicz:79,Jeziorski:94} has been applied to generate potential energy surfaces for scattering cross sections calculations, predictions of spectra and bulk matter properties, as well as the development of force fields for biomolecules (see, e.g., Refs.~\onlinecite{Szalewicz:05,Szalewicz:05b,Szalewicz:12,Jansen:14,Patkowski:20}).

In contrast to the rich toolbox dedicated to single-determinantal wave functions, describing intermolecular interactions in complexes which demand multiconfigurational (MC) wave functions remains a challenge. The multiconfigurational treatment is often mandatory for transition metal complexes, open-shell systems, electronically excited states or systems dominated by static correlation effects. From the standpoint of weak intermolecular forces, proper representation of static correlation warranted by expansion in multiple electron configurations is not sufficient. The main difficulty lies in the recovery of the remaining dynamic correlation both within and between the interacting molecules.
The latter effect, giving rise to the attractive dispersion interaction, poses a particular challenge due to its highly nonlocal and long-ranged nature. Although many multireference methods restoring dynamic correlation effects have been developed, neither has yet managed to combine the accuracy and efficiency required for noncovalent interactions. For instance, the popular multireference configuration interaction (MRCI) approach\cite{Werner:88} and multireference perturbation theories\cite{Andersson:90,Celestino:02} are limited by the lack of triple excitations and truncation of the perturbation series at the second order, respectively, and their application to interactions is not trivial. 
First, MRCI is not size-consistent and requires approximate corrections added a posteriori.~\cite{Langhoff:74,Pople:77} In perturbation theories the fulfilment of the strict separability condition depends on the choice of the zeroth-order Hamiltonian.~\cite{vanDam:99,Rintelman:05}
Second, perturbation theories, including complete active space (CAS) perturbation theory (CASPT2)\cite{Andersson:90} and multireference variants of the M{\o}ller-Plesset perturbation theory,~\cite{Hirao:92} suffer from intruder states\cite{Camacho:10} which have to be removed using one of the available shift techniques.~\cite{Chang:12} 
Intruder states are also a significant problem in the development of multireference coupled-cluster theories,~\cite{Jeziorski:10,Evangelista:18} next to numerical instabilities and algebraic complexity. Single-reference coupled-cluster (CC) approaches introduced by Piecuch and co-workers, e.g.\ the CC($P;Q$) formalism,~\cite{Shen:12a,Shen:12b} may be a viable alternative, as indicated by studies of interactions involving stretched intramonomer covalent bonds.~\cite{Chipman:06,Pastorczak:17} Encouraging results have recently been obtained for strongly correlated interacting systems
from multiconfigurational random phase approximation theory combined with generalized valence bond method.\cite{Pastorczak:17,Pernal:19,Pastorczak:18c}
The multiconfiguration density functional theory (MC DFT)\cite{Stoll:85,Savin:96} methods corrected to include long-range dynamic correlation via perturbation theory\cite{Fromager:10}, the adiabatic connection formalism\cite{Hapka:20} or semi-empirical dispersion models\cite{Stein:20} are also worth mentioning, but their accuracy for noncovalent interations remains to be rigorously investigated.

A separate challenge, parallel to advancing supermolecular methods, is the development of energy decomposition schemes applicable to multireference systems.~\cite{Nine:19} On that front, the SAPT formalism offers several important advantages which make it one of the most widely used and actively developed decomposition approaches.~\cite{Patkowski:20} The interaction energy components in SAPT have a clear physical interpretation and the most accurate variants of the method predict interaction energies closely matching the coupled-cluster singles-and-doubles with perturbative triples [CCSD(T)\cite{Raghavachari:89,Bartlett:90}] results. Compared to the supermolecular approach, SAPT avoids the basis set superposition error, since the interaction energy is computed directly based only on monomer properties. Moreover, the interaction energy represented as a sum of energy contributions is per se size-consistent.

In more than forty years spanning the development of SAPT, applications going beyond the single-reference treatment of the monomers are scarce. Exact full configuration interaction (FCI) wave functions are feasible only for model, few-electron dimers and have been employed in studies of SAPT convergence.~\cite{Korona:99,Patkowski:01,Patkowski:04,Przybytek:04,Szalewicz:05} Reinhardt\cite{Reinhardt:17} used valence bond (VB) wave functions to represent the electrostatic interaction between monomers of a multireference character and proposed an approximate, VB-based approach for dispersion energy calculations. The spin-flip SAPT (SF-SAPT)\cite{Patkowski:18,Waldrop:19} formalism introduced by Patkowski and co-workers opened the possibility to treat multireference, low-spin states based on single reference description of the subsystems. First-order spin-flip exchange energy expressions for high-spin restricted open-shell Hartree-Fock (ROHF) wave function have already been derived and implemented,~\cite{Patkowski:18,Waldrop:19} while extension to the second-order is under way.~\cite{Patkowski:20} 

The purpose of the present paper is to present a complete SAPT formalism applicable to interactions involving multireference systems. First steps in this directions have been already taken. Recently, we have devised multiconfigurational approaches for second-order dispersion\cite{Hapka:19a} and exchange-dispersion\cite{Hapka:19b} energy calculations. In this work we complement these efforts and propose a variant of SAPT based on MC wave functions, which we refer to as SAPT(MC). The method includes energy contributions up to the second-order in the intermolecular interaction operator and can be applied with any wave function model which gives access to one- and two-electron reduced density matrices of the monomers. Following the developments of Ref.~\onlinecite{Hapka:19a} and \onlinecite{Hapka:19b}, the linear response properties required for second-order terms are accessed by solving extended random phase approximation (ERPA)\cite{Pernal:12,Pernal:14b} equations. Both first- and second-order exchange terms are derived assuming the single-exchange approximation,~\cite{Murrell:65} also known as the $S^2$ approximation. We discuss the performance of SAPT(MC) combined either with generalized valence bond perfect pairing (GVB) or complete active space self consistent field (CASSCF) description of the monomers.

This work is organized in five sections. In Section~\ref{sec:theo} we present formulas for first- and second-order energy contributions in the ERPA-based variant of multiconfigurational SAPT. Special attention is given to calculations of induction and dispersion energies for complexes involving electronically excited molecules. Section~\ref{sec:compdet} contains details of our implementation and computations. Results for the model multireference and single-reference dimers are presented in Section~\ref{sec:res}. In Section~\ref{sec:concl} we summarize our findings.

\section{Theory \label{sec:theo}}
Consider a weakly interacting dimer $AB$ which dissociates into monomer $A$ in state $I$ described with the $\ket{\Psi^A_I}$ wave function and monomer $B$ in state $J$ described with the $\ket{\Psi^B_J}$ wave function ($I$ and $J$ refer either to ground or excited states of the monomers). When the unperturbed Hamiltonian is chosen as the sum of Hamiltonians of the isolated monomers, $\hat{H}_0 = \hat{H}_A + \hat{H}_B$, the zeroth-order wave function takes a product form $\ket{\Psi^{(0)}} = \ket{\Psi^A_I\Psi^B_J}$. In this work we assume that $\ket{\Psi^{(0)}}$ is nondegenerate.

The intermolecular interaction operator, $\hat{V}$, represents the perturbation and gathers all Coulombic interactions between electrons and nuclei of the interacting partners
\begin{equation}
\begin{split}
\hat{V} &= \sum_{i \in A} \sum_{j \in B} \frac{1}{r_{ij}} - \sum_{i \in A} \sum_{\beta \in B} \frac{Z_{\beta}}{r_{\beta i}} - \sum_{j \in B} \sum_{\alpha \in A} \frac{Z_{\alpha}}{r_{\alpha j}} \\ 
&+ \sum_{\alpha \in A} \sum_{\beta \in B} \frac{Z_\alpha Z_\beta}{r_{\alpha\beta}}  \\
&= \sum_{i \in A}\sum_{j \in B} \frac{1}{r_{ij}} + \sum_{i \in A} v^B(\mathbf{r}_i) + \sum_{j \in B} v^A(\mathbf{r}_j) + V^{AB} \\
\end{split}
\end{equation}
where $i$ and $j$ run over $N_A$ and $N_B$ electrons in monomers $A$ and $B$, respectively, $v^{A}$ and $v^B$ are one electron potentials, and $V^{AB}$ is the nuclear-nuclear repulsion term.

In the symmetrized Rayleigh-Schr{\"o}dinger (SRS) formulation\cite{Jeziorski:78} of SAPT the interaction energy is expanded with respect to $\hat{V}$ while enforcing the antisymmetry of $\ket{\Psi^A_I}$ and $\ket{\Psi^B_J}$ wave function products. The general expression for energy contribution in the $n$-th order in $\hat{V}$ takes the form
\begin{equation}
\begin{split}
E^{(n)}_{\rm SRS} &= \frac{\braket{\Psi_I^A \Psi_J^B | \hat{V} \hat{\mathcal{A}} \Psi_{\rm RS}^{(n-1)}}}{\braket{\Psi_I^A \Psi_J^B | \hat{\mathcal{A}} \Psi_I^A \Psi_J^B}} \\
&- \frac{\sum_{k=1}^{n-1} E_{\rm SRS}^{(k)}\braket{\Psi_I^A \Psi_J^B | \hat{V} \hat{\mathcal{A}} \Psi_{\rm RS}^{(n-k)}}}{\braket{\Psi_I^A \Psi_J^B | \hat{\mathcal{A}} \Psi_I^A \Psi_J^B}} 
\end{split}
\end{equation}
where $\hat{\mathcal{A}}$ is the antisymmetrizer exchanging electrons between the monomers; $\ket{\Psi^{(n)}_{\rm RS}}$ denotes $n$-th order component of the wave function expansion, which is identical in both SRS and in the conventional Rayleigh-Schr{\"o}dinger (RS) perturbation theory, i.e., the expansion is based only on simple products of zero-order functions. The difference between the SRS and RS energy is defined as the exchange energy. The RS energy contributions are often referred to as polarization components. For convenience, we use the SAPT acronym when referring to SRS.   

The SAPT(MC) formalism presented in this work includes interaction energy components through the second-order in $\hat{V}$
\begin{equation}
\begin{split}
E_{\rm int}^{\rm SAPT} &= E_{\rm elst}^{(1)} + E_{\rm exch}^{(1)} + E_{\rm ind}^{(2)} + E_{\rm exch-ind}^{(2)} \\
&+ E_{\rm disp}^{(2)} + E_{\rm exch-disp}^{(2)}
\end{split}
\label{eq:sapt2}
\end{equation}
where $E^{(1)}_{\rm elst}$ and $E^{(1)}_{\rm exch}$ are first-order electrostatic and exchange energy contributions, respectively, $E^{(2)}_{\rm ind}$ and $E^{(2)}_{\rm exch-ind}$ are the second-order induction and exchange-induction energies, respectively, and $E^{(2)}_{\rm disp}$ and $E^{(2)}_{\rm exch-disp}$  denote the dispersion energy and its exchange counterpart, respectively.

All formulas are in the natural-orbitals (NOs) representation. We use the following index convention: greek $\mu$ and $\nu$ indices denote electronic states of monomers, $p_{\sigma}q_{\sigma}r_{\sigma}s_{\sigma}$ denote natural spinorbitals, while $pqrs$ pertain to natural orbitals denoted by $\varphi(\mathbf{r})$. Throughout the work the NOs are assumed to be real-valued. In the representation of natural spinorbitals the one-electron reduced density matrix (1-RDM) is diagonal
\begin{equation}
\gamma_{pq} = \frac{1}{2} \left\langle \Psi| \hat{a}_{q_{\alpha}}^{\dagger}\hat{a}_{p_{\alpha}} + \hat{a}_{q_{\beta}}^{\dagger}\hat{a}_{p_{\beta}} |\Psi \right\rangle = n_p \delta_{pq}
\end{equation}
where $\left\{ \hat{a}_{p_{\sigma}}^{\dagger}\right\}$ and $\left\{ \hat{a}_{p_{\sigma}}\right\}$ are creation and annihilation operators, respectively, and $n_p$ are natural occupation numbers from the $\braket{0,1}$ range, summing up to half a number of electrons, $\sum_p n_p = N/2$.

All presented SAPT(MC) energy contributions are given in a spin-summed form. The expressions for the polarization energy components are valid for arbitrary spin states of the monomers. The exchange energy contributions are presented assuming singlet spin states of the monomers which implies that $\alpha\alpha$ and $\beta\beta$ blocks of 1-RDM are equal.

\subsection{First-order energy contributions}

The polarization component of the first-order SAPT energy is the electrostatic energy, $E^{(1)}_{\rm elst} = \braket{\Psi^A_I \Psi^B_J | \hat{V} | \Psi^A_I \Psi^B_J}$. This energy contribution expressed in terms of 1-RDMs takes the form
\begin{equation}
\begin{split}
E_{\rm elst}^{(1)} &= 2 \sum_{p \in A} n_p v^B_{pp} + 2 \sum_{q \in B} n_q v^A_{qq} \\ &+ 4 \sum_{\substack{p \in A \\ q \in B}} n_p n_q v_{pq}^{pq} + V^{AB}
\end{split}
\end{equation}
where $v^{A(B)}_{pq} = \braket{\varphi_p|v^{A(B)}|\varphi_q}$ are matix elements of one-electron potentials and $v_{pq}^{rs}$ denote regular two-electron Coulomb integrals $v_{pq}^{rs} = \braket{\varphi_p(\mathbf{r}_1)\varphi_q(\mathbf{r}_2)|r_{12}^{-1}|\varphi_r(\mathbf{r}_1)\varphi_s(\mathbf{r}_2)}$.

Evaluation of the exact expression for the first-order exchange energy
\begin{equation}
E^{(1)}_{\rm exch} = \frac{\braket{\Psi_I^A \Psi_J^B | \hat{V} \hat{\mathcal{A}} \Psi_I^A \Psi_J^B}}{\braket{\Psi_I^A \Psi_J^B | \hat{\mathcal{A}} \Psi_I^A \Psi_J^B}} - E^{(1)}_{\rm elst}
\label{eq:exch1}
\end{equation}
requires access to many-particle density matrices of the monomers. For the Hartree-Fock wave function many-particle density matrices are readily available as antisymmetrized products of the one-particle density matrix.\cite{Jeziorski:76} At the SAPT(DFT) level of theory one uses approximate 1-RDMs of the monomers based on the Kohn-Sham determinants and employs the same exchange expression as in the wave-function SAPT.~\cite{Hesselmann:02a,Misquitta:02}

It is possible to significantly simplify the structure of Eq.~\eqref{eq:exch1} by allowing only for single-exchange of electrons between the monomers in the antisymmetrizer\cite{Murrell:65,Bulski:79} 
\begin{equation}
E^{(1)}_{\rm exch}(S^2) = \braket{\Psi^A_I\Psi^B_J|\left(\hat{V}-E^{(1)}_{\rm elst}\right)\hat{\mathcal{P}}|\Psi^A_I\Psi^B_J}
\end{equation}
where the single-exchange operator $\hat{\mathcal{P}}$ collects all permutations, $\hat{P}_{ij}$, interchanging the coordinates of electrons $i$ and $j$
\begin{equation}
\hat{\mathcal{P}} = -\sum_{i \in A}\sum_{j \in B} \hat{P}_{ij}
\label{eq:P1}
\end{equation}
Neglecting multiple exchange of electrons is known as the $S^2$ approximation and allows one to express the first-order exchange energy using only 1-RMDs and two-electron reduced density matrices (2-RDMs) of the monomers.~\cite{Moszynski:94} Following the density-matrix-based formulation of Ref.~\onlinecite{Moszynski:94} we obtain
\begin{eqnarray}
E_{\rm exch}^{(1)}(S^{2}) & = & 2 \left( E_{\rm elst}^{(1)} - V_{AB} \right) \sum_{\substack{ p\in A \\ q\in B}} n_{p}n_{q} \left( S_p^q \right)^{2} \notag \\ 
&& -2 \sum_{\substack{ p\in A \\ q\in B}} n_{p}n_{q} \left( v_{pq}^{A} S_p^q  + v_{qp}^{B} S_p^q + v_{pq}^{qp} \right) \notag \\
&& -4 \sum_{\substack{ pqr\in A \\ t\in B}} n_{t} \ N_{pqrt}^{A}\ (v_{pr}^{B}\ S_q^t + v_{pq}^{rt}) \notag \\
&& -4 \sum_{\substack{ pqr\in B  \\ t\in A}} n_{t}\ N_{pqrt}^{B}\ (v_{pr}^{A}\ S_q^t + v_{pq}^{rt}) \notag \\
&& -8 \sum_{\substack{ pqr\in A  \\ sut\in B}} \ N_{pqru}^{A}\ N_{tusq}^{B}\ v_{pt}^{rs}
\label{eq:exchs2}
\end{eqnarray}
where $S_p^q = \braket{\varphi_p|\varphi_q}$ denotes the overlap integral, and we have introduced intermediates containing contractions of the  2-RDM, $\Gamma_{p_{\sigma}q_{\sigma'}r_{\sigma''}s_{\sigma'''}} = \braket{\Psi|\hat{a}_{r_{\sigma''}}^{\dagger}\hat{a}_{s_{\sigma'''}}^{\dagger}\hat{a}_{q_{\sigma'}}\hat{a}_{p_{\sigma}}|\Psi}$, with the overlap integrals
\begin{equation}
  N^A_{tuvw} = \sum_{a \in A} \bar{\Gamma}^A_{tuva} S_a^w, \qquad
  N^B_{tuvw} = \sum_{b \in B} \bar{\Gamma}^B_{tuvb} S_w^b
\label{eq:Nint}
\end{equation}
where $\bar{\Gamma}_{pqrs}$ is the spin-summed 2-RDM, $\bar{\Gamma}_{pqrs}= \Gamma_{p_{\alpha}q_{\alpha}r_{\alpha}s_{\alpha}} + \Gamma_{p_{\beta}q_{\alpha}r_{\beta}s_{\alpha}}$. Since we assume monomers in singlet states, the $\beta\beta\beta\beta$+$\alpha\beta\alpha\beta$ block is equal to its $\alpha\alpha\alpha\alpha$+$\beta\alpha\beta\alpha$ counterpart.

\subsection{Second-order energy contributions}

\subsubsection{Transition properties from Extended Random Phase approximation}\label{subsec:transition}

Second-order SAPT energy components may be expressed through transition properties of the interacting monomers. The induction and dispersion energies involve transition energies and one-electron reduced transition densities (1-TRDMs). The SRS components, exchange-induction and exchange-dispersion energies, require both 1-TRDMs and two-electron reduced transition densities (2-TRDMs).

In this work we approximate the transition properties of the interacting monomers by solving the Extended Random Phase Approximation\cite{Rowe:68,Pernal:12,Hapka:19a} equations (independently for each monomer)
\begin{equation}
\left(
\begin{array}
[c]{cc}%
\mathbf{\mathcal{A}} & \mathbf{\mathcal{B}}\\
\mathbf{\mathcal{B}} & \mathbf{\mathcal{A}}%
\end{array}
\right)  \left(
\begin{array}
[c]{c}%
\mathbf{X}_{\nu}\\
\mathbf{Y}_{\nu}
\end{array}
\right)  =\omega_{\nu}\left(
\begin{array}
[c]{cc}%
-\mathbf{\mathcal{N}} & \mathbf{0} \\
\mathbf{0} & \mathbf{\mathcal{N}}
\end{array}
\right)  \left(
\begin{array}
[c]{c}%
\mathbf{X}_{\nu}\\
\mathbf{Y}_{\nu}%
\end{array}
\right)  \ \ \ ,\label{eq:erpa}
\end{equation}
with a diagonal metric matrix
\begin{equation}
\forall_{\substack{p>q\\r>s}} \ \ \ \mathcal{N}_{pq,rs}=(n_{p}-n_{q})\delta_{pr}\delta_{qs}
\end{equation}
The ERPA equations may be formed as a symmetric real eigenproblem using electronic Hessian matrices, $\mathcal{A}+\mathcal{B}$ and $\mathcal{A}-\mathcal{B}$. 
For ground state calculations the Hessian matrices are positive definite (see, e.g., Refs.~\onlinecite{Pastorczak:15} and \onlinecite{Pastorczak:18a} for explicit ERPA equations in the GVB and CAS frameworks, respectively). In the case of excited state wave functions the Hessian matrices may have negative eigenvalues corresponding to de-excitation modes in the ERPA propagator\cite{Golab:83} (see a more detailed discussion in section \ref{sec:ExcSt}).

Apart from transition energies $\omega_\nu$ which correspond to the poles of the ERPA eigenproblem, two quantities that are required in second-order SAPT are 1- and 2-TRDM of the monomers. The 1-TRDM is defined as
\begin{equation}
\gamma_{p_{\sigma}q_{\sigma}}^{\nu}=\left\langle \Psi|\hat{a}_{q_{\sigma}}^{\dagger}\hat{a}_{p_{\sigma}}|\Psi_{\nu} \right\rangle
\end{equation}
Note that for singlet states $\alpha\alpha$ and $\beta\beta$ blocks are equal: $\gamma^{\nu}_{p_{\alpha}q_{\alpha}}$ = $\gamma^{\nu}_{p_{\beta}q_{\beta}}$ = $\gamma^{\nu}_{pq}$. 
The general definition of 2-TRDM reads
\begin{equation}
\Gamma^{\nu}_{p_{\sigma}q_{\sigma'}r_{\sigma''}s_{\sigma'''}}=\braket{\Psi|\hat{a}_{r_{\sigma''}}^{\dagger}\hat{a}_{s_{\sigma'''}}^{\dagger}\hat{a}_{q_{\sigma'}}\hat{a}_{p_{\sigma}}|\Psi_{\nu}}
\end{equation}
The 1-TRDM are expressed through the ERPA eigenvectors as\cite{Pernal:14,Pastorczak:18a}
\begin{align}
\forall_{p>q}\ \ \ \gamma_{qp}^{\nu} &  =(n_{p}-n_{q})\left[ \mathbf{Y}_\nu \right]  _{pq} \label{eq:gam_Y}\\
\forall_{q>p}\ \ \ \gamma_{qp}^{\nu} &  =(n_{p}-n_{q})\left[ \mathbf{X}_\nu \right]_{qp} \label{eq:gam_X}
\end{align}
and the formula for the half of spin-summed 2-TRDM reads\cite{Hapka:19b}
\begin{equation}
\begin{split}
\bar{\Gamma}^{\nu}_{pqrs} &= \Gamma^{\nu}_{p_{\alpha}q_{\alpha}r_{\alpha}s_{\alpha}} + \Gamma^{\nu}_{p_{\beta}q_{\alpha}r_{\beta}s_{\alpha}} \\
& =\sum_{t<p}[\mathbf{X}_\nu]_{pt}\bar{\Gamma}_{tqrs} + \sum_{t<q}[\mathbf{X}_\nu]_{qt}\bar{\Gamma}_{ptrs} \\
          & - \sum_{t>r}[\mathbf{X}_\nu]_{tr}\bar{\Gamma}_{pqts} - \sum_{t>s} [\mathbf{X}_\nu]_{ts}\bar{\Gamma}_{pqrt} \\
          & + \sum_{t>p}[\mathbf{Y}_\nu]_{tp}\bar{\Gamma}_{tqrs} + \sum_{t>q} [\mathbf{Y}_\nu]_{tq}\bar{\Gamma}_{ptrs} \\
          & - \sum_{t<r}[\mathbf{Y}_\nu]_{rt}\bar{\Gamma}_{pqts} - \sum_{t<s} [\mathbf{Y}_\nu]_{st}\bar{\Gamma}_{pqrt} \\
\end{split}
\label{eq:2trdm}
\end{equation}

\subsubsection{Induction and dispersion energies}

The polarization components of SAPT in the second order are the induction and
dispersion energies. The induction energy is given as
\begin{equation}
\begin{split}
E_{\mathrm{ind}}^{(2)}  &=-\sum_{\mu\neq I}\frac{\left\vert \langle{\Psi_{I}^{A}\Psi_{J}^{B}|\hat{V}|\Psi_{\mu}^{A}\Psi_{J}^{B}}\rangle\right\vert^{2}}{\omega_{\mu}^{A}} \\ 
& -\sum_{\nu\neq J}\frac{\left\vert \langle{\Psi_{I}^{A}\Psi_{J}^{B}|\hat{V}|\Psi_{I}^{A}\Psi_{\nu}^{B}}\rangle\right\vert ^{2}}{\omega_{\nu}^{B}} \\
& =E_{\mathrm{ind}}^{(2)}(A\leftarrow B)+E_{\mathrm{ind}}^{(2)}(B\leftarrow A)
\end{split}
\label{eq:e2ind}
\end{equation}
where $\omega_{\mu}^{A}$ ($\omega_{\nu}^{B}$) are transition energies from the state $I$ ($J$ for the monomer $B$) to $\mu$ ($\nu$)
\begin{equation}
\omega_{\mu}^{A}=E_{\mu}^{A}-E_{I}^{A} \label{eq:omega}%
\end{equation}
The $E_{\mathrm{ind}}^{(2)}(A\leftarrow B)$ term arises from the permanent
multipole moments on $B$ changing the wave function of monomer $A$. The
$E_{\mathrm{ind}}^{(2)}(B\leftarrow A)$ term describes a corresponding change in
monomer $B$ due to the perturbing field of $A$.

Eq.~\eqref{eq:e2ind} may be recast using contractions between 1-TRDMs of one monomer and the electrostatic potential of its unperturbed interacting partner, the latter defined as
\begin{equation}
\hat{\Omega}_{B}(\mathbf{r})=v^{B}(\mathbf{r})+\int\frac{\rho^{B}(\mathbf{r}^{\prime})}{\left\vert \mathbf{r}-\mathbf{r}^{\prime}\right\vert }{\rm d} \mathbf{r}^{\prime}
\end{equation}
where $\rho^{B}$ is the one-electron density of the monomer $B$ (analogous expression holds for $\hat{\Omega}_{A}$). The total induction energy formula is now conveniently expressed as
\begin{equation}
\begin{split}
E_{\mathrm{ind}}^{(2)} &= -\sum_{\mu\neq I}\frac{\Big( \sum_{pq\in A} \gamma_{pq}^{A,\mu}\ \Omega_{pq}^{B}\Big)^{2}}{\omega_{\mu}^{A}} \\
& -\sum_{\nu\neq J}\frac{ \Big( \sum_{rs\in B} \gamma_{rs}^{B,\nu}\ \Omega_{rs}^{A} \Big)^{2}}{\omega_{\nu}^{B}} \\
\end{split}
\end{equation}
where $\Omega_{pq}=\braket{\varphi_p|\hat{\Omega}|\varphi_q}$. 

In the ERPA-approximation the spin-summed formula for $E_{\mathrm{ind}}^{(2)}$ takes the form
\begin{equation}
\begin{split}
E_{\mathrm{ind}}^{(2)} &= -4\sum_{\mu\in A}\frac{\Big( \sum_{p>q\in A} [\mathbf{Z}^A_\mu]_{pq} \Omega_{pq}^{B} \Big)^2}{\omega_{\mu}^{A}} \\
& -4 \sum_{\nu\in B}\frac{\Big( \sum_{r>s\in B} [\mathbf{Z}^B_\nu]_{rs}  \Omega_{rs}^{A} \Big)^2}{\omega_{\nu}^{B}} \\
\label{eq:indErpa}
\end{split}
\end{equation}
where
\begin{equation}
\begin{split}
\forall_{p>q \in A} \;\; [\mathbf{Z}^A_\mu]_{pq} &= \mathcal{N}_{pq} \Big( \left[  \mathbf{Y}_{\mu}^{A}\right]_{pq}-\left[\mathbf{X}_{\mu}^{A}\right]_{pq} \Big) \\
\forall_{r>s \in B} \;\; [\mathbf{Z}^B_\nu]_{rs} &= \mathcal{N}_{rs} \Big(  \left[  \mathbf{Y}_{\nu}^{B}\right]_{rs}-\left[\mathbf{X}_{\nu}^{B}\right]_{rs} \Big) \\
\end{split}
\end{equation}

The pertinent expression for the dispersion energy is\cite{McWeeny:59,Jaszunski:85}
\begin{equation}
E_{\mathrm{disp}}^{(2)}=-\sum_{\mu\neq I,\nu\neq J}\frac{\left(
\sum_{\substack{pq\in A\\rs\in B}}\gamma_{pq}^{A,\mu}\gamma_{rs}^{B,\nu}%
v_{pr}^{qs}\right)  ^{2}}{\omega_{\mu}^{A}+\omega_{\nu}^{B}} \label{edisp2}%
\end{equation}
which in the ERPA form reads\cite{Hapka:19a}
\begin{equation}
E_{\mathrm{disp}}^{(2)}=-16\sum_{\mu\in A,\nu\in B}\frac{\left(
\sum_{\substack{p>q\in A\\r>s\in B}} [\mathbf{Z}^A_\mu]_{pq} [\mathbf{Z}^B_\nu]_{rs} v_{pr}^{qs}\right)  ^{2}}{\omega_{\mu}^{A}
+\omega_{\nu}^{B}}\ \ \ \label{eq:dispErpa}
\end{equation}

\subsubsection{Excited state case: explicit contributions to dispersion and
induction energies from de-excitations \label{sec:ExcSt}}

Consider a dimer $A_{I}B_{0}$ in an excited state, which dissociates into a monomer $A$ in the state $I>0$ denoted in this section as $A_{I}$ (for simplicity it is assumed that states of $A$ are not degenerate) and a monomer $B$ in the ground state, denoted as $B_{0}$. While all transition energies, cf.\ Eq.~(\ref{eq:omega}), corresponding to $B_{0}$ are positive
\begin{equation}
\forall_{\nu>0}\ \ \ \omega_{\nu}^{B_{0}}>0
\end{equation}
for the monomer $A$ they take either negative or positive values for transitions to states lower or higher than $I$, respectively
\begin{align}
\forall_{\mu<I}\ \ \ \omega_{\mu}^{A_{I}} &  <0\\
\forall_{\mu>I}\ \ \ \omega_{\mu}^{A_{I}} &  >0
\end{align}
Let us rewrite the dispersion energy expression, Eq.~(\ref{edisp2}), in a form in which we explicitly isolate terms involving negative transitions (de-excitations)
\begin{equation}
\begin{split}
E_{\mathrm{disp}}^{(2)}(A_{I}B_{0}) &= E_{\mathrm{disp}_{+}}^{(2)}(A_{I}B_{0}) \\ 
&- \sum_{\mu<I,\nu>0}\frac{\left(  \sum_{\substack{pq\in A\\rs\in B}}\gamma_{pq}^{A_{I},\mu}\gamma_{rs}^{B_{0},\nu}v_{pr}^{qs}\right)^{2}}{-\left\vert \omega_{\mu}^{A_{I}}\right\vert +\left\vert \omega_{\nu}^{B_{0}}\right\vert }
\end{split}
\label{eq:dispexcit}
\end{equation}
where by $E_{\mathrm{disp}_{+}}^{(2)}(A_{I}B_{0})$ we denote the dispersion energy arising from the positive part of the monomer $A$ linear response  function spectrum
\begin{equation}
E_{\mathrm{disp}_{+}}^{(2)}(A_{I}B_{0})=-\sum_{\mu>I,\nu>0}\frac{\left(
\sum_{\substack{pq\in A\\rs\in B}}\gamma_{pq}^{A_{I},\mu}\gamma_{rs}%
^{B_{0},\nu}v_{pr}^{qs}\right)  ^{2}}{\left\vert \omega_{\mu}^{A_{I}%
}\right\vert +\left\vert \omega_{\nu}^{B_{0}}\right\vert }\label{eq:disp+}%
\end{equation}
In Eqs.~(\ref{eq:dispexcit}) and (\ref{eq:disp+}) signs of the transition energies in the denominators are written explicitly.

Approximate methods which are based on single-excitation operators, and for which the linear response is directly related to an orbital Hessian matrix,
are likely to miss de-excitations in the linear response function computed for the excited state of interest.~\cite{Pernal:21} Consequently, the second term in Eq.~\eqref{eq:dispexcit} would not be accounted for. Since this term involves transitions to the low-lying states, it is anticipated to give a non-negligible contribution to the dispersion energy.

A viable way to account for the de-excitations from the $A_I$ state in dispersion energy calculations is by considering linear response properties of states $J$ lower than $I$. After exploiting the relations connecting response properties of the states $I$ and $J$
\begin{equation}
\begin{split}
\omega_{J}^{A_{I}} & = E_{J}^{A}-E_{I}^{A} \\ &= -\left(  E_{I}^{A}-E_{J} ^{A}\right) = -\omega_{I}^{A_{J}} \\
\end{split}
\end{equation}
\begin{equation}
\begin{split}
\gamma_{pq}^{A_{I},J} & =\left\langle \Psi_{I}|\hat{a}_{q}^{\dagger} \hat{a}_{p} | \Psi_{J}\right\rangle \\
&= \left\langle \Psi_{J}|\hat{a}_{p}^{\dagger} \hat{a}_{q}|\Psi_{I}\right\rangle ^{\ast} = \left( \gamma_{qp}^{A_{J},I} \right)^{\ast}
\end{split}
\end{equation}
one immediately writes the $\mu=J$ component of Eq.~\eqref{eq:dispexcit} as \begin{equation}
-\sum_{\nu>0}\frac{\left(  \sum_{\substack{pq\in A_{J}\\rs\in B_{0}}%
}\gamma_{pq}^{A_{J},I}\gamma_{rs}^{B_{0},\nu}v_{pr}^{qs}\right)  ^{2}%
}{-\left\vert \omega_{I}^{A_{J}}\right\vert +\left\vert \omega_{\nu}^{B_{0}%
}\right\vert }\equiv\varepsilon_{\mathrm{disp}}^{I\rightarrow J}(A_{J}B_{0})
\end{equation}
The dispersion energy for the $A_{I}B_{0}$ dimer can now be written as
\begin{equation}
E_{\mathrm{disp}}^{(2)}(A_{I}B_{0})=E_{\mathrm{disp}_{+}}^{(2)}(A_{I}%
B_{0})+\sum_{J=0}^{I-1}\varepsilon_{\mathrm{disp}}^{I\rightarrow J}(A_{J}%
B_{0})\label{eq:disp_corr}%
\end{equation}
It should be emphasized that Eq.~\eqref{eq:disp_corr} is fully equivalent to Eq.~\eqref{edisp2} if exact response properties are employed. The crucial
difference between the expressions in Eqs.~\eqref{eq:dispexcit} and
\eqref{eq:disp_corr} is that in the former contributions to the
dispersion energy from negative excitations follow from the linear response of the state $A_{I}$, while in the latter they are obtained from the response of states $A_{J}$ which are lower in energy than $A_{I}$.

The ERPA model applied to excited-state reference wave function either completely misses negative excitations or reproduces them with poor accuracy. As a result, ERPA-approximated dispersion energy, Eq.~\eqref{eq:dispErpa}, computed for the excited-state dimer $A_{I}B_{0}$ will lack important contributions from dexcitations. The way around this problem is to employ the alternative formula for the dispersion energy presented in Eq.~\eqref{eq:disp_corr} in the ERPA\ approximation. This requires computing the $E_{\mathrm{disp}_{+}}^{(2)}(A_{I}B_{0})$ term according to Eq.~\eqref{eq:dispErpa} and expressing the
approximated $\varepsilon_{\mathrm{disp}}^{I\rightarrow J}(A_{J}B_{0})$ terms through ERPA transition properties
\begin{equation}
\begin{split}
& \varepsilon_{\mathrm{disp}}^{I\rightarrow J}(A_{J}B_{0}) = \\ 
&-16\sum_{\nu>0} \frac{\left(  \sum_{\substack{p>q\in A_{J}\\r>s\in B_{0}}}\left[ \mathbf{Z}_{I}^{A_{J}}\right]_{pq}\left[  \mathbf{Z}_{\nu}^{B_{0}}\right]_{rs}v_{pr}^{qs}\right)  ^{2}}{-\left\vert \omega_{I}^{A_{J}}\right\vert + \left\vert \omega_{\nu}^{B_{0}}\right\vert }
\end{split}
\label{eq: epsdisp}
\end{equation}
The $\mathbf{Z}_{I}^{A_{J}}$ and $\omega_{I}^{A_{J}}$ are the $I$-th
eigenvector and eigenvalue, respectively, of the ERPA\ equations solved for
the monomer $A$ in the $J$-th state
\begin{align}
\mathbf{Z}_{I}^{A_{J}}  & =\mathbf{Z}_{I}^{A_{J}}(\Psi_{J}^{A})\\
\omega_{I}^{A_{J}}  & =\omega_{I}^{A_{J}}(\Psi_{J}^{A})
\end{align}
To reiterate, the negative-energy transition
$I\rightarrow J$, which is either absent or erroneous in ERPA, is easily accessed through a positive-energy transition $J\rightarrow I$ computation carried out for the states $J<I$. A similar approach has recently been applied to improve the description of the correlation energy for excited states within the adiabatic connection ERPA method.~\cite{Pernal:21} Notice that for the lowest excited states, which are usually of interest, the $\varepsilon^{I \to J}_{\mathrm{disp}}(A_{J}B_{0})$ terms have a negative sign, but could in principle be positive for highly excited state $I$.

The second-order induction energy for a dimer in the excited state, obtained
with the ERPA\ approximation, Eq.~\eqref{eq:indErpa}, has to be corrected for the missing de-excitations in an analogous manner
\begin{equation}
E_{\mathrm{ind}}^{(2)}(A_{I}B_{0})=E_{\mathrm{ind}_{+}}^{(2)}(A_{I}B_{0}%
)+\sum_{J=0}^{I-1}\varepsilon^{I \to J}_{\mathrm{ind}}(A_{J}B_{0})
\label{eq:ind_corr}
\end{equation}
The $E_{\mathrm{ind}_{+}}^{(2)}(A_{I}B_{0})$ term is obtained from Eq.~\eqref{eq:indErpa} where the sum with respect to $\mu$ runs through positive
transitions ($\omega_{\mu}^{A}>0$). The $\varepsilon^{I \to J}_{\mathrm{ind}}
(A_{J}B_{0})$ terms are given as
\begin{equation}
\varepsilon^{I \to J}_{\mathrm{ind}}(A_{J}B_{0})= 4\frac{\Big(\sum_{p>q\in A_{J}%
}\left[  \mathbf{Z}_{I}^{A_{J}}\right]_{pq}\Omega_{pq}^{B_{0}}\Big)^{2}%
}{|\omega_{I}^{A_{J}}|}%
\label{eq: epsind}
\end{equation}
and follow from solving ERPA\ equations for the monomer $A$ in states lower than $I$ (from the ground state, $J=0$, up to $J=I-1$). Evidently,
contributions to the induction energy from negative excitations always take a positive sign.

\subsubsection{Second-order exchange energy contributions}

We begin with the general expressions for the second-order induction and exchange-dispersion energies in the $S^2$ approximation\cite{Chalbie:77a,Chalbie:77b}
\begin{equation}
\begin{split}
E^{(2)}_{\rm exch-ind}(S^2) &= \braket{\Psi_I^A\Psi_J^B|\hat{V}\hat{\mathcal{P}} \Psi_{\rm ind}^{(1)}} \\
& - E^{(1)}_{\rm elst} \braket{\Psi_I^A\Psi_J^B|\hat{\mathcal{P}} \Psi_{\rm ind}^{(1)}} \\
& - E_{\rm ind}^{(2)}\braket{\Psi_I^A\Psi_J^B|\hat{\mathcal{P}}|\Psi_I^A\Psi_J^B} \\
E^{(2)}_{\rm exch-disp}(S^2) &= \langle \Psi_I^A\Psi_J^B | \hat{V}\hat{\mathcal{P}} \Psi^{(1)}_{\rm disp} \rangle \\
& - E^{(1)}_{\rm elst}\langle  \Psi_I^A\Psi_J^B | \hat{\mathcal{P}} \Psi^{(1)}_{\rm disp} \rangle \\
& - E^{(2)}_{\rm disp}\langle  \Psi_I^A\Psi_J^B | \hat{\mathcal{P}} \Psi_I^A\Psi_J^B \rangle \\
\end{split}
\end{equation}
where $\ket{\Psi_{\rm ind}^{(1)}}$ and $\ket{\Psi_{\rm disp}^{(1)}}$ are the first order induction and dispersion wave functions, respectively
\begin{equation}
\begin{split}
\Psi_{\rm ind}^{(1)} &= -\sum_{\mu \neq I}\frac{\ket{\Psi_{\mu}^A\Psi_J^B}\braket{\Psi_{\mu}^A\Psi_J^B|\hat{V}\Psi^A_I\Psi^B_J}}{E^A_\mu-E^A_I} \\
&- \sum_{\nu \neq J} \frac{\ket{\Psi_I^A\Psi_{\nu}^B}\braket{\Psi_I^A\Psi_{\nu}^B | \hat{V}\Psi^A_I\Psi^B_J}}{E^B_{\nu}-E^B_J} \\
\Psi^{(1)}_{\rm disp} &= -\sum_{\mu \neq I, \nu\neq J } \frac{\ket{\Psi^A_\mu\Psi^B_\nu}\braket{\Psi^A_\mu\Psi^B_\nu|\hat{V}| \Psi^A_I\Psi^B_J }}{E_\mu^A + E_\nu^B - E_I^A - E_J^B} \\
\end{split}
\end{equation}

First calculations of second-order exchange contributions in the single exchange approximation for many-electron systems were perfomed by Cha{\l}asi{\'n}ski and Jeziorski.~\cite{Chalbie:77b} The authors derived general expressions in the form of a many-orbital cluster expansion based on the induction and dispersion pair functions. Expressions in terms of one-electron orbital basis set were given in Ref.~\onlinecite{Rybak:91} for the exchange-dispersion energy and in Ref.~\onlinecite{Jeziorski:93} for the exchange-induction contributions. During the development of the SAPT(CCSD) approach, Korona presented the density-matrix formulation of both second-order exchange components.~\cite{Korona:08,Korona:09}

For ground-state single-determinant wave function or Kohn-Sham determinant is it possible to calculate second-order exchange terms through all orders in the intermolecular overlap, as proven by Sch{\"a}ffer and Jansen.~\cite{Schaffer:12,Schaffer:13} Recently, Waldrop and Patkowski have derived expressions for the third-order exchange-induction.~\cite{Waldrop:21} 

The exchange-induction energy written in terms of density matrices and transition energies reads (the $S^2$ notation is dropped for convenience)
\begin{strip}
\begin{equation}
\begin{split}
E_{\rm exch-ind}^{(2)}(A \leftarrow B) & =-\sum_{\mu\neq I, \nu=0} \frac{\left\langle
\left(  \tilde{\upsilon}(\mathbf{r},\mathbf{r}^{\prime})+\frac{V_{AB}}%
{N_{A}N_{B}}\right)  \gamma_{\rm int}^{A,\mu}(\mathbf{x},\mathbf{x}^{\prime
})\right\rangle \left\langle \gamma^{A,\mu}(\mathbf{x},\mathbf{x})\left\vert
\mathbf{r}-\mathbf{r}^{\prime}\right\vert ^{-1}\gamma^{B}(\mathbf{x}
^{\prime},\mathbf{x}^{\prime})\right\rangle }{\omega_{\mu}^{A}} \nonumber \\
& - E_{\rm elst}^{(1)}\sum_{\mu\neq I,\nu=0} \frac{\left\langle \gamma^{A,\mu
}(\mathbf{x},\mathbf{x}^{\prime}) \gamma^{B}(\mathbf{x}^{\prime}
,\mathbf{x})\right\rangle \left\langle \gamma^{A,\mu}(\mathbf{x}
,\mathbf{x})\left\vert \mathbf{r}-\mathbf{r}^{\prime}\right\vert^{-1}
\gamma^{B}(\mathbf{x}^{\prime},\mathbf{x}^{\prime})\right\rangle}
{\omega_{\mu}^{A}} \nonumber\\
& + E_{\rm ind}^{(2)}\left\langle \gamma^{A}(\mathbf{x},\mathbf{x}^{\prime}%
)\gamma^{B}(\mathbf{x}^{\prime},\mathbf{x})\right\rangle
\end{split}
\end{equation}
\end{strip}
where $\tilde{\upsilon}(\mathbf{r},\mathbf{r}^{\prime})$ is the generalized interaction potential
\begin{equation}
\forall_{\substack{r\in A\\r^{\prime}\in B}} \ \ \tilde{\upsilon}%
(\mathbf{r},\mathbf{r}^{\prime})=\left\vert \mathbf{r}-\mathbf{r}^{\prime
}\right\vert ^{-1}+\frac{\upsilon_{B}(\mathbf{r})}{N_{B}}+\frac{\upsilon
_{A}(\mathbf{r}^{\prime})}{N_{A}}
\label{eq:vgen}
\end{equation}
and $\gamma_{\rm  int}^{A,\mu}$ stands for the interaction density matrix\cite{Moszynski:94,Korona:08}
\begin{align}
\small
& \gamma_{\rm int}^{A,\mu}(\mathbf{x},\mathbf{x}^{\prime}) = \\ 
& -\gamma^{A,\mu}(\mathbf{x}^{\prime},\mathbf{x})\gamma^{B}(\mathbf{x},\mathbf{x}^{\prime}) \nonumber \\
& -\int\Gamma^{A,\mu}(\mathbf{x},\mathbf{x}^{\prime};\mathbf{x},\mathbf{x}^{\prime\prime})\gamma^{B}(\mathbf{x}^{\prime\prime},\mathbf{x}^{\prime})d\mathbf{x}^{\prime\prime} \nonumber \\
& -\int\gamma^{A,\mu} (\mathbf{x}^{\prime\prime},\mathbf{x})\Gamma^{B}(\mathbf{x}^{\prime},\mathbf{x};\mathbf{x}^{\prime},\mathbf{x}^{\prime\prime}){\rm d}\mathbf{x}
^{\prime\prime} \nonumber \\
&  -\int\int\Gamma^{A,\mu}(\mathbf{x},\mathbf{x}^{\prime\prime\prime
};\mathbf{x},\mathbf{x}^{\prime\prime})\Gamma^{B}(\mathbf{x}^{\prime
},\mathbf{x}^{\prime\prime};\mathbf{x}^{\prime},\mathbf{x}^{\prime\prime
\prime}) {\rm d}\mathbf{x}^{\prime\prime}{\rm d}\mathbf{x}^{\prime\prime\prime}
\end{align}
The 1- and 2-TRDMs in the position representation are defined as 
\begin{equation}
\small
\begin{split}
& \gamma^{A,\mu}(\mathbf{x}_{1},\mathbf{x}_{1}^{\prime}) = \\
&N_{A} \int \Psi_{I}^{A}(\mathbf{x}_{1},\ldots)^{\ast}\Psi_{\mu}^{A}(\mathbf{x}_{1}^{\prime},\ldots){\rm d}\mathbf{x}_{2} {\rm d}\mathbf{x}_{3}\ldots {\rm d}\mathbf{x}_{N_{A}}
\end{split}
\end{equation} 
and
\begin{equation}
\small
\begin{split}
& \Gamma^{A,\mu}(\mathbf{x}_{1},\mathbf{x}_{2};\mathbf{x}_{1}^{\prime
},\mathbf{x}_{2}^{\prime}) = \\
&N_{A}(N_{A}-1) \int \Psi_{I}^{A}(\mathbf{x}_{1},\mathbf{x}_{2},...)^{\ast}\Psi_{\mu}^{A}(\mathbf{x}%
_{1}^{\prime},\mathbf{x}_{2}^{\prime},...) {\rm d}\mathbf{x}_{3} ...
 {\rm d}\mathbf{x}_{N_{A}}
\end{split}
\label{eq:trdm1}
\end{equation}
The pertinent expressions for the $E^{(2)}_{\rm exch-ind}(B \leftarrow A)$ component follow by interchanging $A$ and $B$ indices.

In Ref.~\onlinecite{Hapka:19b} we have derived the density-matrix formula for the exchange-dispersion energy based on transition properties in the ERPA framework. The corresponding expression for the exchange-induction energy component takes the form 
\begin{equation}
\begin{split}
& E^{(2)}_{\rm exch-ind} = \\ 
&4 \sum_{\mu \in A} \frac{V_\mu^A u_{\mu}^A}{\omega^A_{\mu}} + 4 \sum_{\nu \in B} \frac{V_\nu^B u_\nu^B}{\omega^B_\nu}  \\
&- 4 \left( E^{(1)}_{\rm elst}-V^{AB} \right) \left(\sum_{\mu \in A} \frac{u^A_\mu t^A_\mu}{\omega^A_\mu} + \sum_{\nu \in B} \frac{u^B_\nu t^B_\nu}{\omega^B_\nu} \right) \\ 
&+ 2E^{(2)}_{\rm ind}\sum_{\substack{p \in A \\ q \in B}} n_p n_q (S_p^q)^2 \\
\end{split}
\label{eq:e2exi}
\end{equation}
The intermediates in Eq.~\eqref{eq:e2exi} read
\begin{equation}
\begin{split}
u_\mu^A = \sum_{p>q \in A} \left( [\textbf{Y}_{\mu}^A]_{pq}-[\textbf{X}_{\mu}^A]_{pq} \right) (n_p-n_q) \Omega^B_{pq} \\
u_\nu^B = \sum_{r>s \in B} \left( [\textbf{Y}_{\nu}^B]_{rs}-[\textbf{X}_{\nu}^B]_{rs} \right) (n_r-n_s) \Omega^A_{rs}
\end{split}
\end{equation}
\begin{equation}
\begin{split}
t_\mu^A = \sum_{\substack{p>q \in A \\ r \in B}} \left( [\textbf{Y}_{\mu}^A]_{pq}-[\textbf{X}_{\mu}^A]_{pq} \right) (n_p-n_q)n_r S_p^r S_q^r \\
t_\nu^B = \sum_{\substack{r>s \in B \\ p \in A}} \left( [\textbf{Y}_{\nu}^B]_{rs}-[\textbf{X}_{\nu}^B]_{rs} \right) (n_r-n_s)n_p S_r^p S_s^p \\
\end{split}
\end{equation}
and 
\begin{equation}
\begin{split}
V_\mu^A = &\sum_{\substack{pq \in A \\ r \in B }} n_r \gamma^{A,\mu}_{pq} \tilde{v}_{qp}^{rr} 
  + \sum_{\substack{pqaa' \in A \\ r \in B }} n_r \bar{\Gamma}^{A,\mu}_{pqaa'} S_{a'}^r \tilde{v}_{pq}^{ar} \\
& + \sum_{\substack{pq \in A \\ rsbb' \in B }} \gamma^{A,\mu}_{pq} \bar{\Gamma}^{B}_{rsbb'} S_p^{b'}\tilde{v}_{qb}^{sr} \\
& + \sum_{\substack{pqaa' \in A \\ rsbb' \in B }} \bar{\Gamma}^{A,\mu}_{pqaa'} \bar{\Gamma}^{B}_{rsbb'} S_q^{b'} S_{a'}^s \tilde{v}_{pr}^{ab}
\end{split}
\label{eq:Va}
\end{equation}
\begin{equation}
\begin{split}
V_\nu^B &= \sum_{\substack{p \in A \\ rs \in B }} n_p \gamma^{B,\nu}_{rs}\tilde{v}_{pp}^{rs} + \sum_{\substack{p \in A \\ rsbb' \in B }} n_p \bar{\Gamma}^{B,\nu}_{rsbb'} S_p^{b'}\tilde{v}_{pb}^{sr} \\
& +\sum_{\substack{pqaa' \in A \\ rs \in B }} \gamma^{B,\nu}_{rs} \bar{\Gamma}^{A}_{pqaa'} S_{a'}^r \tilde{v}_{pq}^{as} \\
& +\sum_{\substack{pqaa' \in A \\ rsbb' \in B }} \bar{\Gamma}^{A}_{pqaa'} \bar{\Gamma}^{B,\nu}_{rsbb'} S_q^{b'} S_{a'}^s \tilde{v}_{pr}^{ab}
\end{split}
\label{eq:Vb}
\end{equation}
where the effective two-electron potential, Eq.~\eqref{eq:vgen} in the matrix representation is
\begin{equation}
\begin{split}
\tilde{v}_{pq}^{rs} &= \bigg \langle \varphi_p(\mathbf{r})\varphi_q(\mathbf{r}') | \tilde{v}(\mathbf{r},\mathbf{r}') | \varphi_r(\mathbf{r})\varphi_s(\mathbf{r}') \bigg \rangle \\
&= v_{pq}^{rs} + N_B^{-1}\braket{\varphi_p|v^B|\varphi_r}\big(\delta_{qs}\delta_{X_q X_s} + S_q^s (1-\delta_{X_q X_s})\big) \\
&+ N_A^{-1}\braket{\varphi_q|v^A|\varphi_s}\big( \delta_{pr}\delta_{X_p X_r} + S_p^r(1-\delta_{X_p X_r}) \big) \ \ \ ,
\end{split}
\end{equation}
(a $\varphi_p$ orbital may belong either to monomer $X_p = A$ or $X_p = B$).

When 1- and 2-TRDMs in Eqs.~\eqref{eq:Va}-\eqref{eq:Vb} are expanded according to Eqs.~\eqref{eq:gam_Y}-\eqref{eq:gam_X} and Eq.~\eqref{eq:2trdm}, respectively, one arrives at the matrix representation of the $V^A_\mu$ and $V^B_\nu$ terms
\begin{equation}
\begin{split}
V_\mu^A = &-\sum_{p>q \in A} [ \mathbf{X}^A_{\mu}]_{pq} \Big[ (n_p - n_q) \sum_{r \in B} n_r \tilde{v}_{pq}^{rr} \\ 
& + \sum_{r \in B} n_r P^A_{pqrr} + (n_p-n_q) Q^B_{pq} - R^{AB}_{pq} \Big] \\
& - \sum_{p>q \in A} [ \mathbf{Y}^A_{\mu}]_{pq} \Big[-(n_p - n_q) \sum_{r \in B} n_r \tilde{v}_{pq}^{rr} \\
& + \sum_{r \in B} n_r P^A_{qprr} - (n_p-n_q) Q^B_{qp} - R^{AB}_{qp} \Big] \\
\end{split}
\label{eq:va}
\end{equation}

\begin{equation}
\begin{split}
V_\nu^B = &-\sum_{r>s \in B} [\mathbf{X}^B_{\nu}]_{rs} \Big[ (n_r-n_s) \sum_{p \in A} n_p \tilde{v}_{pp}^{rs} \\
& + \sum_{p \in A} n_p P^B_{pprs} + (n_r-n_s) Q^A_{rs} - R^{BA}_{rs} \Big] \\
& - \sum_{r>s \in B} [\mathbf{Y}^B_{\nu}]_{rs} \Big[-(n_r-n_s) \sum_{p \in A} n_p \tilde{v}_{pp}^{rs} \\
& + \sum_{p \in A} n_p P^B_{ppsr} - (n_r-n_s) Q^A_{sr} - R^{BA}_{sr} \Big]
\end{split}
\label{eq:vb}
\end{equation}
where
\begin{equation}
 \label{eq:pa}
\begin{split}
 P^A_{pqrs} &= \sum_{aa' \in A} \big( 
 N^A_{aa'pr} \tilde{v}_{qa'}^{as}
 - N^A_{qa'ar} \tilde{v}_{pa'}^{as}
 - N^A_{aqa'r} \tilde{v}_{a'p}^{as}\big) \\
 & + O^A_{ps} S_q^r \\
 P^B_{pqrs} &= \sum_{bb' \in B} \big(
  N^B_{bb'rp} \tilde{v}_{qb}^{b's}
- N^B_{sb'bp} \tilde{v}_{qb}^{b'r} 
- N^B_{bsb'p} \tilde{v}_{qb}^{rb'}\big) \\
& + O^B_{rq} S_p^s \\
Q^A_{rs} &= \sum_{a \in A} O^A_{as} S_a^r \\
Q^B_{pq} &= \sum_{b \in B} O^B_{bq} S_p^b \\
%
 R^{AB}_{pq} &= \sum_{\substack{a \in A \\ bb' \in B}} T_{qabb'} \tilde{v}_{pb'}^{ab} - T_{apbb'} \tilde{v}_{ab}^{qb'} + W_{aqb'b} S_p^{b'} S_a^b \\
& - W_{pab'b} S_a^{b'} S_q^b \\
 R^{BA}_{rs} &= \sum_{\substack{aa' \in A \\ b \in B}} T_{aa'sb} \tilde{v}_{ra'}^{ba} - T_{aa'br} \tilde{v}_{ba}^{sa'} + W_{a'abs} S_{a'}^r S_a^b \\
& - W_{a'arb} S_{a'}^b S_a^s
\end{split}
\end{equation}
with $N^A$ and $N^B$ intermediates given in Eq.~\eqref{eq:Nint}, and the remaining intermediates defined as
\begin{equation}
\begin{split}
O^A_{tu} &= \sum_{aa'a'' \in A} \bar{\Gamma}^A_{taa'a''} \tilde{v}_{aa'}^{a''u} \\
O^B_{tu} &= \sum_{bb'b'' \in B} \bar{\Gamma}^B_{tbb'b''} \tilde{v}_{ub''}^{b'b}
\end{split}
\end{equation}
\begin{equation}
\begin{split}
T_{tuvw} &= \sum_{\substack{aa' \in A \\ bb' \in B }} \bar{\Gamma}^A_{taua'}\bar{\Gamma}^B_{bvb'w} S_{a'}^{b} S_{a}^{b'} \\
         & = \sum_{bb' \in B} \bigg(\sum_{aa' \in A} \bar{\Gamma}^A_{taua'} S_{a}^{b'} S_{a'}^{b}\bigg)\bar{\Gamma}^B_{bvb'w}
\end{split}
\label{eq:intT}	
\end{equation}
\begin{equation}
\begin{split}
W_{tuvw} &= \sum_{\substack{aa' \in A \\ bb' \in B }} \bar{\Gamma}^A_{taua'}\bar{\Gamma}^B_{bvb'w} \tilde{v}_{ab}^{a'b'} \\
         &= \sum_{bb' \in B} \bigg(\sum_{aa' \in A} \bar{\Gamma}^A_{taua'}\tilde{v}_{ab}^{a'b'}\bigg)\bar{\Gamma}^B_{bvb'w}
\end{split}
\label{eq:intW}
\end{equation}

Both induction and exchange-induction terms in SAPT are routinely calculated in the coupled approximation,~\cite{Jaszunski:80} so that the response of monomer orbitals due to the perturbation field of its interacting partner is accounted for. The uncoupled approach, which neglects the influence of the perturbing field, is used in calculations of the dispersion and exchange dispersion energies in wave-function SAPT\cite{Jeziorski:94} including the popular SAPT0 model.~\cite{Hohenstein:10,Parker:14} In both SAPT(DFT) and SAPT(CCSD) the coupled level of theory has been shown to give highly accurate second-order energy contributions.~\cite{Misquitta:02,Hesselmann:02b,Hesselmann:03,Misquitta:03,Korona:08a,Korona:08,Korona:09}

Evaluation of the exchange-induction energy requires construction of the $\mathbf{T}$ and $\mathbf{W}$ intermediates, Eq.~\eqref{eq:intT} and Eq.~\eqref{eq:intW}, respectively, which has the $n_{\rm OCC}^6$ scaling ($n_{\rm OCC}$ are orbitals with non-zero occupancy). Since the 2-RDM matrix elements factorize unless all four indices correspond to fractionally occupied orbitals, the formal scaling with the sixth power is only with respect to the number of such orbitals. 
In comparison, the exchange-dispersion energy is more expensive, as it requires steps with a $n_{\rm OCC}^3 n_{\rm SEC}^3$ scaling ($n_{\rm SEC} = M_{s_2} + M_{s_3}$).~\cite{Hapka:19b} It should also be noted that the bottleneck step in evaluation of the first-order exchange energy, Eq.~\eqref{eq:exchs2}, engages three four-index quantities (the $\mathbf{N}^A$, $\mathbf{N}^B$ intermediates and integrals) which amounts to $M_{s_2}^6$ scaling. Note that for GVB the 2-RDMs factorize also in the active block\cite{Pernal:12} which results in identical scaling as in the SAPT(HF) method.

Recently, we have demonstrated that the uncoupled approximation in the ERPA framework combined either with CASSCF or GVB description of the monomers leads to a poor quality of the second-order dispersion energy.~\cite{Hapka:19a,Hapka:19b} A more accurate dispersion energy is obtained if the monomer response properties are expanded up to the first order in the coupling parameter, which we refer to as the semicoupled approximation.~\cite{Hapka:19a} The fully coupled ERPA scheme gives best results for both dispersion and exchange-dispersion energies. In this work all second-order energy components were obtained with the coupled approximation.

The multiconfigurational SAPT method, comprising first- and second-order energy components, is based on the chosen wave function theory applied to description of monomers. It is important to notice that computation of all SAPT terms requires only knowledge of the corresponding one- and two-electron reduced density matrices of monomers. In the rest of this work we use the notation SAPT(MC) for the proposed method, where MC indicates the underlying multiconfigurational wave function model employed to obtain reduced density matrices. The results will be presented for two multiconfigurational wave funcitons: CASSCF and GVB approximations. For comparison, we also include the SAPT results following from the single-determinantal description of monomers, denoted as SAPT(HF).

\section{Computational details \label{sec:compdet}}

The ERPA equations applied to GVB or CAS wave functions require dividing the orbital space of each monomer into three disjoint subsets referred to as $s_1$, $s_2$ and $s_3$. The cardinalities of the subsets are represented by the $M_{s_1}$, $M_{s_2}$ and $M_{s_3}$ notation. For wave functions of the CAS type the $s_1$ set contains all inactive orbitals, whereas $s_2$ and $s_3$ correspond to the active and virtual orbitals, respectively. When ERPA is applied with the GVB reference, the $s_1$ set is defined as all orbitals which occupation numbers fulfil the $n_p > 0.992$ condition. The $s_2$ set includes all active orbitals, i.e., strongly occupied orbitals with occupation numbers $0.992 \geq n_p \geq 0.5$ and their weakly occupied partners from the same geminal. The remaining orbitals are grouped in the $s_3$ set. The $p, q$ indices of the $[\mathbf{X}_\nu]_{pq}$ and $[\mathbf{Y}_\nu]_{pq}$ vectors span the following range
\begin{equation}
\begin{matrix}
p \in s_2 \wedge q \in s_1 \\
p \in s_3 \wedge q \in s_1 \\
p \in s_2 \wedge q \in s_2 \\
p \in s_3 \wedge q \in s_2 \\
\end{matrix}
\label{eq:ss}
\end{equation}
(analogous range is assumed for the $pq$ and $rs$ indices of $\mathcal{A}_{pq,rs} \pm \mathcal{B}_{pq,rs}$).

In ERPA the presence of degeneracies and near-degeneracies in the $p \in s_2 \wedge q \in s_2$ space, cf.\ Eq.~\eqref{eq:ss}, may lead to numerical instabilities. To circumvent this, we discarded pairs of orbitals [in practical terms, it means discarding corresponding rows and columns in the ERPA matrices, see Eq.~\eqref{eq:erpa}] applying the $|n_p - n_q|/n_p < 10^{-2}$ condition for the GVB wave function and $|n_p - n_q|/n_p < 10^{-6}$ for the CAS wave function.

The results obtained with CASSCF and GVB treatment of the monomers are denoted as SAPT(CAS), and SAPT(GVB), respectively. Pertinent calculations were performed in the locally developed code. The necessary integrals, 1- and 2-RDMs for CASSCF wave functions were obtained from a developer version of the \textsc{Molpro} program.~\cite{Molpro:12} The GVB calculations were carried out in the locally modified Dalton program\cite{Dalton:13} and interfaced with our code. The MP2 natural orbitals were used as the starting guess in both CASSCF and GVB calculations. 

For the \ce{H2}$\cdots$\ce{H2} dimer, discussed in Section~\ref{sec:h2h2}, we carried out reference calculations exact up to second-order in SAPT, using an in-house code developed for interactions between two-electron monomers and based on direct projection onto irreducible representations of the symmetric $S_4$ group.~\cite{Korona:97} The pertinent results are denoted as SAPT(FCI) in this work.

The augmented correlation-consistent orbital basis sets of double- and triple-zeta quality (aug-cc-pV$X$Z, $X=$ D,T)\cite{Dunning:89,Kendall:92} were employed throughout the work. Monomer calculations were carried out in the dimer-centered basis set.

In Section~\ref{sec:sref} we present results of SAPT(GVB) and SAPT(CAS) calculations for benchmark data set of noncovalently bound complexes introduced by Korona\cite{Korona:13} which we refer to as the TK21 data set. The accuracy of individual SAPT energy components and interaction energies is verified against the SAPT(CCSD) benchmark. 
All CCSD calculations were performed with frozen core electrons. The SAPT(HF), SAPT(DFT) and SAPT2+(CCD) results are also reported. The exchange-correlation PBE0\cite{Perdew:96,Adamo:99} functional employed in SAPT(DFT) was asymptotically-corrected using the GRAC scheme\cite{Gruning:2001} applied with the experimental values of the ionization potentials. The SAPT(HF), SAPT(DFT) and SAPT(CCSD) calculations were performed in \textsc{Molpro}.~\cite{Molpro:12} The SAPT2+(CCD) results were obtained with the Psi4\cite{Psi4:17} program. In the latter variant of SAPT, the interaction energy is represented as
\begin{equation}
\begin{split}
& E_{\rm int}^{\rm SAPT2+(CCD)} = E^{(10)}_{\rm elst} + E^{(12)}_{\rm elst,resp} \\ 
& + E^{(10)}_{\rm exch} + E^{(11)}_{\rm exch} + E^{(12)}_{\rm exch} \\
& + E^{(20)}_{\rm ind,resp} + E^{(22)}_{\rm ind} + E^{(20)}_{\rm exch-ind,resp} + E^{(22)}_{\rm exch-ind} \\ 
&+ E^{(2)}_{\rm disp,CCD} + E^{(20)}_{\rm exch-disp} \\
\end{split}
\label{eq:sapt2pccd}
\end{equation}
where the ($ij$) superscript refers to the $i$th- and $j$th-order expansion in the intermolecular interaction operator and intramolecular correlation operator, respectively; the energy terms marked with the ``resp'' index account for the orbital relaxation effects. Except for the $E^{(2)}_{\rm disp,CCD}$ term, the interaction energy components grouped in Eq.~\eqref{eq:sapt2pccd} are identical to the SAPT2\cite{Hohenstein:10b} approach. The ``+(CCD)'' notation indicates that the dispersion energy is obtained in the coupled pair approximation including noniterative contributions from single and triple excitations, here referred to as CCD+ST(CCD)\cite{Williams:95,Parrish:13} approach.

The accuracy of SAPT interaction energies discussed in Section~\ref{sec:sref} is verified against counterpoise-corrected~\cite{Boys:70} (CP) supermolecular CCSD(T) results. To this end, we approximate higher-order induction contributions at the Hartree-Fock level of theory\cite{Jeziorska:87,Moszynski:96}
\begin{equation}
\label{eq:delta}
\begin{split}
\delta_{\rm HF} &= E_{\rm int}^{\rm HF} \\ 
& - \left( E^{(10)}_{\rm elst} + E^{(10)}_{\rm exch} + E^{(20)}_{\rm ind,resp} + E^{(20)}_{\rm exch-ind,resp} \right)
\end{split}
\end{equation}
where $E_{\rm int}^{\rm HF}$ is the supermolecular Hartree-Fock interaction energy. The $\delta_{\rm HF}$ component was added to the SAPT interaction energy provided that the ratio of the sum of the induction and exchange-induction energies to the total interaction energy was larger than 12.5\%, in agreement with the criterion selected in Ref.~\onlinecite{Taylor:16}. Note that in Section~\ref{sec:sref} error statistics for total interaction energies are reported for the S$_2$ subset of the TK21 dataset which excludes six largest dimers (see Refs.~\onlinecite{Korona:13} and \onlinecite{Hapka:20m}). 
 
As an additional test, we performed SAPT calculations for the A24 dataset\cite{Rezac:13} of \v{R}ez\'{a}\v{c} and Hobza. Since we observed the same qualitative trends as in the TK21 case, results for the A24 dataset are given in the Supporting Information.

\section{Results \label{sec:res}}

\subsection{Multi-reference ground-state system: H$_2$-H$_2$ \label{sec:h2h2}}

We begin the analysis of multiconfigurational SAPT with a model \ce{H2}$\cdots$\ce{H2} dimer. We monitor the change of the interaction energy upon bond dissociation in one of the hydrogen molecules. A quantitative description of this system is challenging as it has to capture the balance between long-range dynamic correlation and increasing nondynamic correlation effects.~\cite{Pastorczak:17,Brzek:19} 

\begin{sidewaysfigure*}
\includegraphics[width=\textwidth]{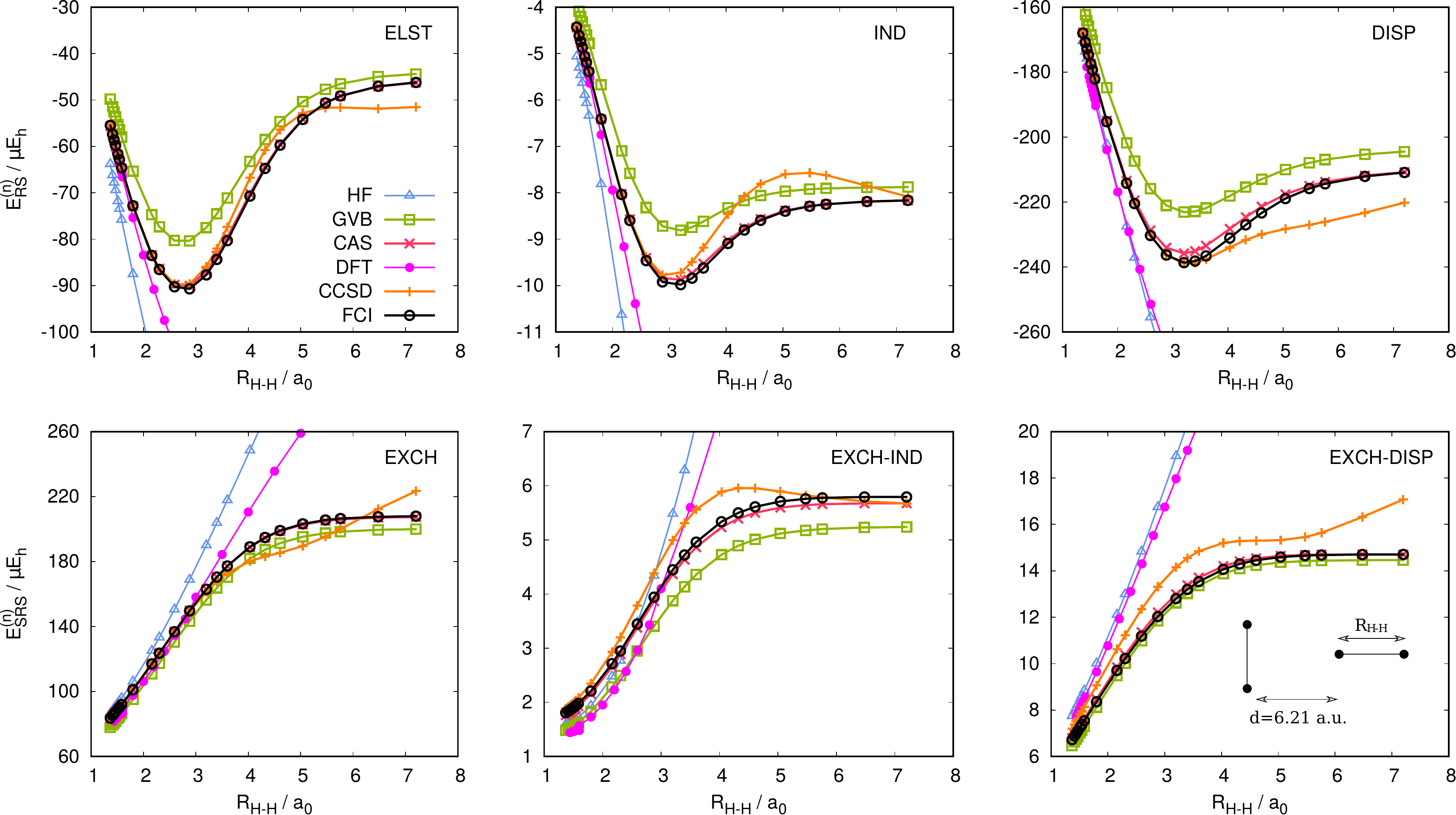}
\caption{SAPT energy contribution at the Hartree-Fock (HF), GVB, CASSCF (CAS), CCSD, DFT and FCI levels of theory for the \ce{H2}$\cdots$\ce{H2} dimer in the T-shaped configuration. The intermolecular distance is fixed at 6.21$\,a_0$. In one of the monomers the $R_{\rm H-H}$ distance is varied from 1.37 to 7.20$\,a_0$ while in the other monomer the H--H bond is kept at a fixed distance of 1.44$\,a_0$. The basis set is aug-cc-pVTZ.}
\label{fig:h2}
\end{sidewaysfigure*}

We examine the T-shaped structure of the \ce{H2}$\cdots$\ce{H2} complex in which one the covalent H-H bonds is stretched from \SI{1.37}{\bohr} to \SI{7.2}{\bohr} (see Figure~\ref{fig:h2} for a detailed description). In SAPT(CAS) calculations each monomer is described with a CAS(2,5) wave function. Note that for two-electron monomers SAPT(GVB) is equivalent to SAPT(CAS) based on CAS(2,2) wave functions.

SAPT schemes based either on Hartree-Fock or Kohn-Sham description of the monomers fail to predict the behavior of individual interaction energy components as the \ce{H-H} bond is elongated and the complex gains a multireference character (Figure~\ref{fig:h2}). The SAPT(CSSD) approach initially remains in excellent agreement with the SAPT(FCI) benchmark. The largest relative percent errors in SAPT(CCSD) near the equilibrium geometry ($R_{\rm H-H} =$ \SI{1.41}{\bohr}) occur for the exchange-induction and exchange-dispersion energies which are overestimated by ca.~$4$\% and $7$\%, respectively. These discrepancies in the single-reference regime result from exclusion of certain cumulant contributions in the second-order exchange expressions.~\cite{Korona:08,Korona:09} After the \ce{H-H} bond length exceeds \SI{3.0}{\bohr}, the XCCSD-3 approximation underlying SAPT(CCSD)\cite{Korona:06a,Korona:06b} starts to break down which translates into qualitative errors in all interaction energy components.~\cite{Pastorczak:17}

\begin{figure*}
\includegraphics[width=0.75\textwidth]{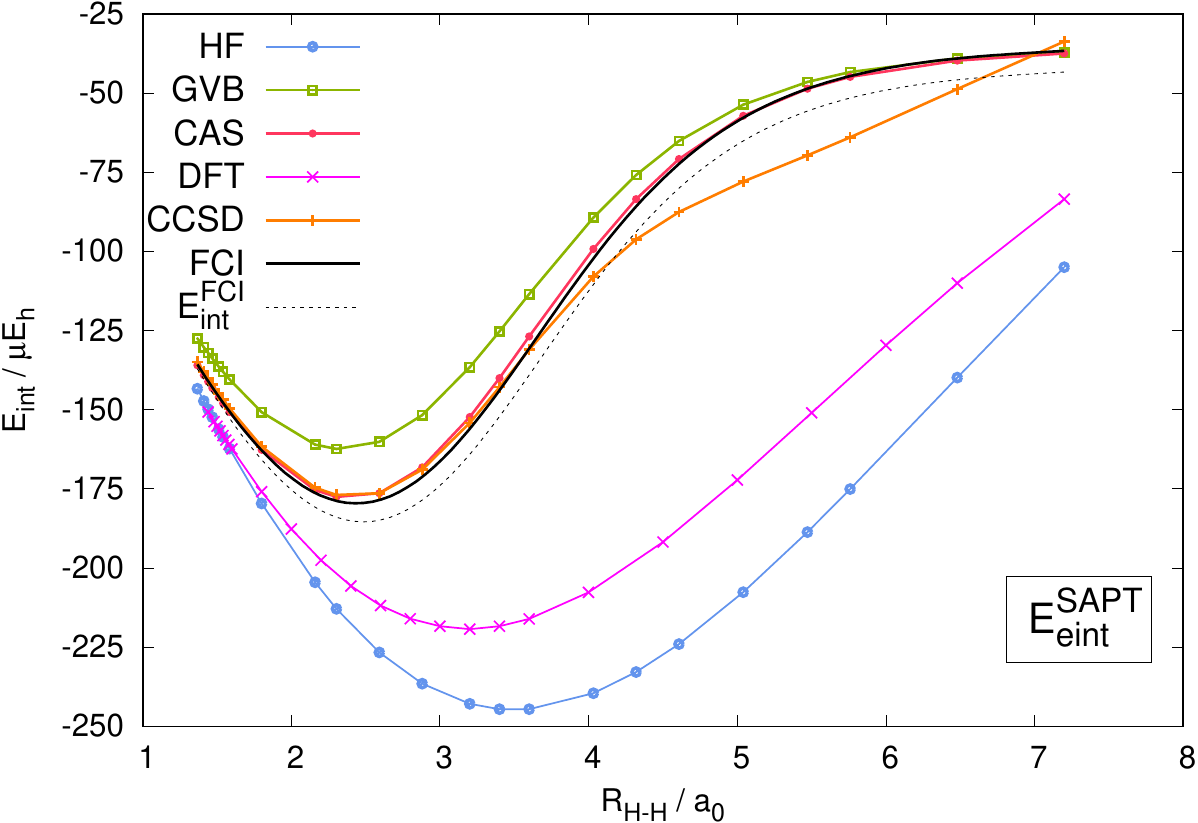}
\caption{SAPT interaction energy for the \ce{H2}$\cdots$\ce{H2} dimer in the T-shaped configuration (see Figure~\ref{fig:h2} for geometry description). $E_{\rm int}^{\rm FCI}$ denotes the supermolecular FCI interaction energy. The basis set is aug-cc-pVTZ.}
\label{fig:h2int}
\end{figure*}

Both SAPT(GVB) and SAPT(CAS) predict the correct shape of the interaction energy curves (Figures~\ref{fig:h2}-\ref{fig:h2int}). The GVB-based variant systematically underestimates the magnitude of all SAPT contributions and SAPT interaction energy. The exchange-induction energy deviates most from the benchmark with relative percent errors in the $10-20$\% range. Errors for the remaining components stay below $12$\% near the equilibrium and the accuracy improves together with the increasing share of the nondynamic correlation in the system (see also Tables S1-S3 in the Supporting Information). SAPT(CAS) is more accurate---errors with respect to the SAPT(FCI) benchmark do not exceed $3$\% not only in individual components,  but also in the total interaction energy. The error of the SAPT(FCI) interaction energy with respect to supermolecular FCI (denoted as $E_{\rm int}^{\rm FCI}$ in Figure~\ref{fig:h2int}) increases from $1.1$\% in the equilibrium geometry to $15$\% at $R_{\rm H-H} =$\SI{7.2}{\bohr}.

As we discussed in Ref.~\onlinecite{Hapka:19a}, further extension of the active space in SAPT(CAS) is of little benefit for this system. Instead, to reach higher accuracy one needs to move beyond the ERPA scheme and solve full linear response equations, i.e., include response not only from the orbitals, but also the wave function expansion coefficients.

\subsection{Excited-state system: C$_2$H$_4^*$-Ar \label{sec:arc2h4}}

\begin{figure*}
\includegraphics[scale=0.68]{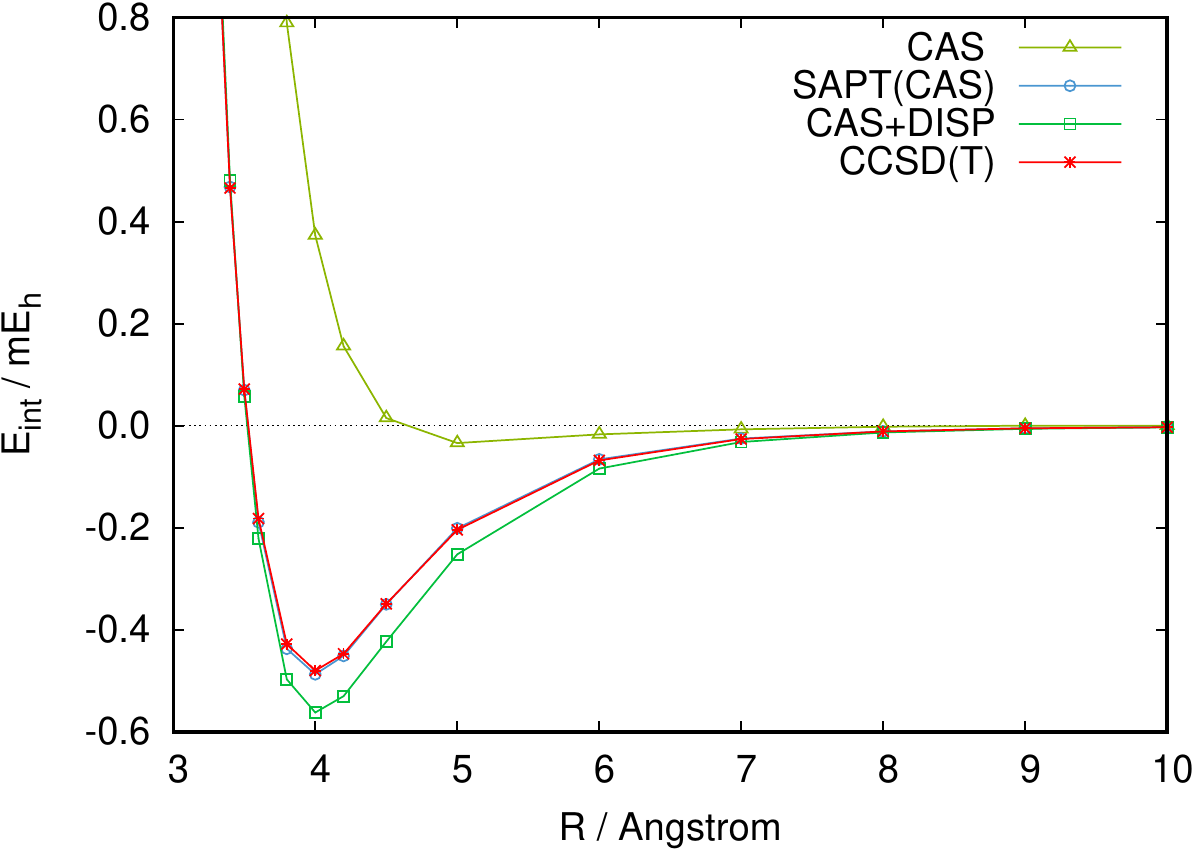}
\includegraphics[scale=0.68]{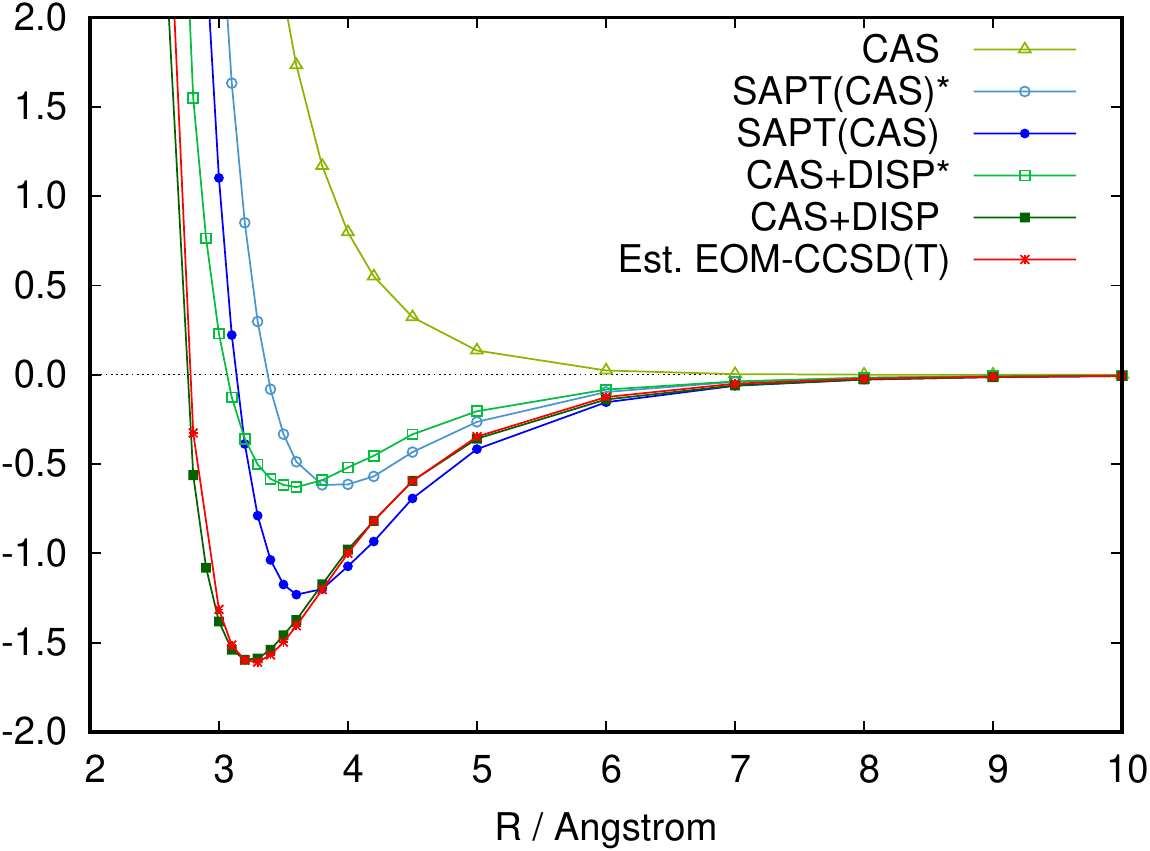}
\caption{Interaction energy curves for the \ce{C2H4}$\cdots$\ce{Ar} dimer in the ground state (left) and in the $\pi \rightarrow \pi^*$ state (right). The basis set is aug-cc-pVTZ.}
\label{fig:arc2h4}
\end{figure*}

In this section we present SAPT(CAS) calculations for the \ce{C2H4}$\cdots$Ar dimer in which the ethylene molecule is either in the ground or electronically excited state. We focus on the singlet excitation of a valence character with the largest contribution from the $\pi \rightarrow \pi^{\ast }$ transition.~\cite{Besley:14}
The \ce{C2H4}$\cdots$\ce{Ar} complex is kept in the $C_{2v}$ symmetry with the Ar atom located on the axis perpendicular to the \ce{C2H4} plane and bisecting its C-C bond (see also Table~S4 in the Supporting Information for geometry of the \ce{C2H4} molecule). The interaction energy curves presented in Figure~\ref{fig:arc2h4} and Table~\ref{tab:sapt} are obtained by varying the distance $R$ between the Ar atom and the center of the mass of ethylene.
Counterpoise correction~\cite{Boys:70} has been applied to all supermolecular interaction energies presented in this section.

To access the excited states wave functions of both the dimer and the ethylene molecule, we carried out three-state state-averaged CAS calculations (SA-CAS) in the CAS(2,3) active space, i.e., two active electrons distributed on $\pi$, $\sigma$, and $\pi ^{\ast}$ active orbitals. In these calculations the targeted $\pi \rightarrow \pi^*$ state is the third state in the SA ensemble. Note that in both SAPT(CAS) and supermolecular CASSCF calculations the Ar atom is represented with a single determinant. 

For ground state calculations we used supermolecular CCSD(T) results as benchmark. To obtain reference values for excited states, we adopted the procedure of Ref.~\onlinecite{Nakai:12} which combines the CCSD(T) description of the ground state with excitation energies calculated at the EOM-CCSD\cite{Monkhorst:77} level of theory
\begin{equation}
\begin{split}
& E_{\mathrm{int}}^{\text{Est.\ EOM-CCSD(T)}}(\mathrm{C}_2\mathrm{H}_4^{\ast }\cdots\mathrm{Ar}) = \\
& E_{\mathrm{int}}^{\text{CCSD(T)}}(\mathrm{C}_2\mathrm{H}_4\cdots\mathrm{Ar}) \notag +\omega_{\mathrm{EOM-CCSD}}(\mathrm{C}_{2}\mathrm{H}_{4}^{\ast }\cdots\mathrm{Ar}) \\
& - \omega_{\mathrm{EOM-CCSD}}(\mathrm{C}_{2}\mathrm{H}_{4}^{\ast })
\label{eq:eom}
\end{split}
\end{equation}
where the asterisk indicates a molecule in excited state, $\omega _{\mathrm{EOM-CCSD}}$ denotes the pertinent EOM-CCSD
excitation energy and $E_{\mathrm{int}}^{\text{CCSD(T)}}$ is a ground state interaction energy.

In Figure~\ref{fig:arc2h4} we compare SAPT(CAS) interaction energy curves with supermolecular CAS(2,3) results and a coupled-cluster benchmark. As it has been rigorously shown in Ref.~\onlinecite{Hapka:20m}, supermolecular CAS interaction energy misses dispersion contributions if active orbitals are assigned only to one monomer, which is the case here.
The CAS+DISP curves in Figure~\ref{fig:arc2h4} represent CAS interaction energy supplemented with the dispersion component taken from SAPT(CAS) calculations, $E^{(2)}_{\rm DISP}=E_{\rm disp}^{(2)}+E_{\rm exch-disp}^{(2)}$. For the $\pi \rightarrow \pi^*$ state 
both SAPT(CAS)\ and CAS+DISP interaction energies  were computed by explicitly accounting for the de-excitation-energy terms according to Eq.~\eqref{eq:disp_corr} and Eq.~\eqref{eq:ind_corr}. The SAPT(CAS)$^{\ast}$ and CAS+DISP$^{\ast}$ curves were obtained by neglecting the $\varepsilon_{\mathrm{disp}}^{I\rightarrow J}$ and $\varepsilon_{\mathrm{ind}}^{I\rightarrow J}$ terms.

Inspection of Figure~\ref{fig:arc2h4} and Table~\ref{tab:sapt} 
shows that the \ce{C2H4}$\cdots$Ar complex in the
ground state is bound by the dispersion forces. The CAS interaction curve is mainly repulsive and features only a shallow minimum located at ca.~\SI{5.0}{\angstrom} and \SI{0.03}{\milli\hartree} deep. Addition of the dispersion energy in CAS+DISP builds up a van der Waals minimum \SI{0.56}{\milli\hartree} deep localized at \SI{4.0}{\angstrom}, which is in reasonable agreement with the CCSD(T) reference ($0.48$~m$E_h$ at $R_{\rm eq} = 4.0$~\AA).
The performance of SAPT(CAS) is excellent. The total SAPT(CAS) interaction energy at the optimal monomer separation is equal to \SI{-0.49}{\milli\hartree} and the entire interaction curve almost coincides with the benchmark. The dispersion energy is clearly the dominating attractive contribution amounting to \SI{-1.03}{\milli\hartree} in the minimum (see Table~\ref{tab:sapt}).

\begin{table}
\centering
\caption{Components of interaction energy for the \ce{C2H4}$\cdots$\ce{Ar} dimer in the ground (g.s.) and excited states. $E_{\rm int}^{\rm SAPT}$ and $E_{\rm int}^{\rm SAPT^*}$ are the total interaction energies
with and without, respectively, the inclusion of contributions from the negative-energy transitions
$\varepsilon_{\mathrm{disp/ind}}^{I\rightarrow J}$ defined in Eqs~(\ref{eq: epsdisp}) and (\ref{eq: epsind}).
All values in m$E_{\rm h}$.}
{\renewcommand{\arraystretch}{1.4}
\begin{tabular}{l S S S}
 \hline\hline
 & \multicolumn{1}{c}{g.s.} & \multicolumn{1}{c}{$\pi \rightarrow \pi^*$} \\
\multicolumn{1}{c}{Component} & \multicolumn{1}{c}{$R$=4.00 \si{\angstrom}} & \multicolumn{1}{c}{$R$=3.30 \si{\angstrom}} \\ \hline
 $E_{\rm elst}^{(1)}$        & -0.257 & -2.486 \\
 $E_{\rm exch}^{(1)}$        &  0.723 &	6.054 \\
 $E_{\rm ind_+}^{(2)}$       & -0.474 & -3.987 \\
 $\sum_J\varepsilon_{\mathrm{ind}}^{I\rightarrow J}$&0.000 & 0.000 \\
 $E_{\rm exch-ind}^{(2)}$    & 0.457  &  4.434 \\
 $E_{\rm disp_+}^{(2)}$      &-1.026  & -4.561 \\
 $\sum_J\varepsilon_{\mathrm{disp}}^{I\rightarrow J}$ &	0.000 & -1.087 \\
 $E_{\rm exch-disp}^{(2)}$   & 0.090  &  0.844 \\
 \hline
 $E_{\rm int}^{\rm SAPT^*}$  & -0.487 & 0.298 \\ $E_{\rm int}^{\rm SAPT} $   & -0.487 &-0.789 \\ \hline\hline
 \end{tabular}
}
\label{tab:sapt}
\end{table}

Computation of the second-order SAPT components in the proposed
SAPT(CAS) approach involves solving the ERPA equations. When the monomer reduced density matrices entering ERPA equations correspond to an unstable CAS solution, either near-instabilities or instabilities may occur in the linear response.
In general, the SA-CAS\ calculation in a small active space bears the risk that wave functions describing higher excited states are not stable
in ERPA.
This is what we have encountered for the SA-CAS $\pi
\rightarrow \pi ^{\ast }$ state of the studied \ce{C2H4}$\cdots$\ce{Ar} dimer (see Figures S1-S3). To avoid instabilities in the ERPA equations, which manifest in discontinuous interaction energy curves, we applied a three-point cubic extrapolation of second-order energy contributions based on the Dyall partitioning of the monomer Hamiltonian\cite{Dyall:95,Rosta:02} and expansion of the ERPA response properties in the coupling parameter.~\cite{Pernal:18,Hapka:19a} The details on the cubic extrapolation model are provided in the Supporting Information.

The interaction energy curves for the $\pi \rightarrow \pi^{\ast}$ state are shown in Figure~\ref{fig:arc2h4}. At the CASSCF level of theory the interaction has a purely repulsive character. The CAS+DISP model gives a binding curve which remains in excellent agreement with the coupled-cluster reference. The employed CC method [Eq.~\eqref{eq:eom}] predicts a \SI{1.61}{\milli\hartree} deep minimum at the intermonomer separation of \SI{3.3}{\angstrom}. The CAS+DISP minimum occurs at a slightly shorter distance of \SI{3.2}{\angstrom} and is \SI{1.60}{\milli\hartree} deep. Note that the nearly perfect agreement with CC has to rest partially on error cancellation, since CAS+DISP neglects contributions from negative excitations in the second-order exchange-dispersion energy (only $\varepsilon_{\rm disp}^{I \to J}$ terms are included in the model). The interaction energy curve from SAPT(CAS) calculations deviates from both CAS+DISP and CC results at the intermediate and short range. SAPT(CAS) localizes the minimum at \SI{3.6}{\angstrom} and underbinds by as much as \SI{0.8}{\milli\hartree} compared to the CC reference (Table~\ref{tab:sapt}). The large discrepancy between second-order SAPT and the hybrid CAS+DISP approach reflects that higher-order induction terms, present in the supermolecular CAS and absent in SAPT, become important already for the low lying $\pi \rightarrow \pi^{\ast}$ valence state. 

Contributions from the negative-energy transitions in the linear response are essential for a quantitative description of the C$_2$H$_4^*\cdots$\ce{Ar} interaction. Neglecting the $\varepsilon^{I \to J}$ terms in SAPT reduces the well depth by a factor of two, cf.\ SAPT(CAS)* results in Figure~\ref{fig:arc2h4}. Similarly, comparing CAS+DISP with CAS+DISP* reveals that a good agreement of CAS+DISP with the coupled-cluster reference is possible only after inclusion of the de-excitation part of the spectrum. The observed energy lowering comes solely from the $\varepsilon_{\rm disp}^{I \to J}$ terms, as the induction counterparts vanish due to symmetry (Table 1). In the van der Waals minimum, the two dispersion terms [$\varepsilon_{\rm disp}^{2 \to 0}$ and $\varepsilon_{\rm disp}^{2 \to 1}$, cf.~Eq.~\eqref{eq:disp_corr}] sum up to \SI{-1.9}{\milli\hartree} which is a sizeable effect considering that positive-energy transitions amount to \SI{-4.6}{\milli\hartree}.

\subsection{Single-reference systems \label{sec:sref}}

In this section we analyze the performance of the multiconfigurational SAPT schemes for many-electron dimers of the TK21 dataset of Korona\cite{Korona:13} against benchmark SAPT(CCSD) results. Additionally, we present SAPT(PBE0) and SAPT2+(CCD) results. Although TK21 includes systems governed by the dynamic rather than static correlation effects, our aim is to determine the level accuracy which could be expected of the studied mulitconfigurational SAPT if applied to multireference systems. Note that in all SAPT calculations the exchange terms were obtained in the $S^2$ approximation. The first-order exchange and second-order exchange-induction contributions in SAPT(CCSD) include the cumulant contributions.\cite{Korona:08b,Korona:08}

\begin{figure*}
\includegraphics[width=\textwidth]{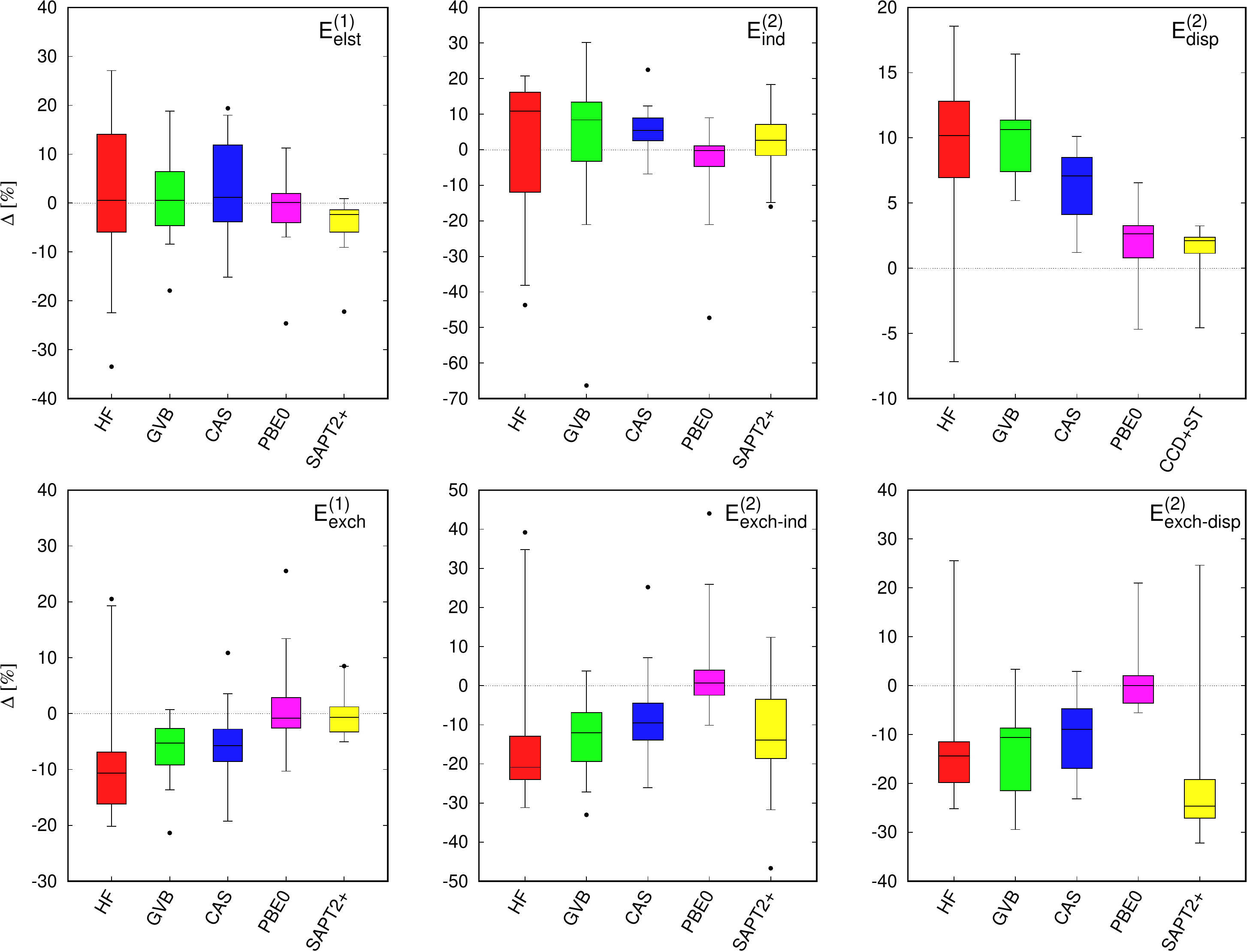}
\caption{Box plots of relative percent errors ($\Delta$) in the polarization SAPT components (top: electrostatic, induction and dispersion energies) and in the exchange SAPT components (bottom: exchange, exchange-induction and exchange-dispersion, all in the $S^2$ approximation) for dimers of the TK21 data set. HF, GVB and CAS denote wave function description of the monomers. PBE0 stands for the asymptotically-corrected PBE0 functional. Relative percent errors are calculated according to the $\Delta_i = \frac{E_{i}-E_{i, {\rm ref}}}{|E_{i, {\rm ref}}|} \cdot 100\%$ formula. Errors are given with respect to the SAPT(CCSD) reference. Errors for the $E^{(2)}_{\rm exch-disp}$ energy are reported for the S$_2$ subset of TK21. The box and outer fences encompass 50\% and 95\% of the distribution, respectively.}
\label{fig:tk21}
\end{figure*}

Figure~\ref{fig:tk21} shows relative percent errors of the individual SAPT energy components with respect to the SAPT(CCSD) reference (see also Tables~S8-S11 in the Supporting Information). Let us begin with the first-order energy terms. Both SAPT(GVB) and SAPT(CAS) recover the electrostatic and exchange energies with similar accuracy---the mean absolute errors ($\overline{\Delta}_{\rm abs}$) for these contributions fall in the 6-8\% range. In contrast to the electrostatic energy, the first-order exchange is systematically underestimated with mean errors of $-6.4$\% and $-5.4$\% obtained with GVB and CAS wavefunctions, respectively. 
SAPT(GVB) affords smaller spread of errors compared to SAPT(CAS), in particular for the $E^{(1)}_{\rm elst}$ component. The multireference treatment of the monomers constitutes an improvement over the Hartree-Fock (single-determinantal) description, (the $\overline{\Delta}_{\rm abs}$ values from the Hartree-Fock-based SAPT calculations amount to ca.\ 13\% for both components) but it remains inferior to both SAPT(DFT) and SAPT2+(CCD).

The second-order SAPT energy contributions obtained with the SAPT(CAS) variant are consistently more accurate than their SAPT(GVB) counterparts. In the TK21 dataset the largest difference occurs for the induction energy, where SAPT(GVB) deviates from the benchmark by $14.6$\% and $20.1$\% in terms of $\overline{\Delta}_{\rm abs}$ and standard deviation, respectively, whereas the respective errors for SAPT(CAS) amount to $7.8$\% and $9.5$\%. This confirms that the CAS-based ERPA provides better approximation for both transition density matrices and transition energies than when GVB density matrices are used.~\cite{Hapka:19a,Hapka:19b} Similar as in the first-order, the polarization terms ($E^{(2)}_{\rm ind}$ and $E^{(2)}_{\rm disp}$) from multiconfigurational SAPT compare favorably with SAPT(HF), but do not match the quality of SAPT(DFT) or SAPT2+(CCD) results. The discrepancy is more pronounced for the dispersion energy where mean absolute errors from CCD+ST(CCD) and SAPT(DFT) calculations are equal to $2.2$\% and $2.9$\%, respectively, compared to $6.4$\% obtained with SAPT(CAS) and $9.9$\% at the SAPT(GVB) level of theory.

The second-order exchange-induction and exchange-dispersion energies are more challenging than the polarization terms. SAPT(DFT) performs best recovering the $E^{(2)}_{\rm exch-ind}$ and $E^{(2)}_{\rm exch-disp}$ contributions with the $\overline{\Delta}_{\rm abs}$ values of $7.5$\% and $4.3$\%, respectively. Both SAPT(GVB) and SAPT(CAS) tend to underestimate second-order exchange ($\overline{\Delta}_{\rm abs}$ values fall in the 10-14\% range). It is worthwhile to note that in the SAPT2+(CCD) scheme the exchange-dispersion energy is calculated using the uncoupled formula which leads to the mean absolute error as large as $24$\% with respect to the coupled SAPT(CCSD) reference.

In Table~\ref{tab:eint} we examine the accuracy of total SAPT interaction energies with respect to the SAPT(CCSD) reference evaluated for the S$_2$ subset of the TK21 dataset. Error statistics is given in terms of the mean error $\overline{\Delta}$, the standard deviation $\sigma$, the mean absolute error $\overline{\Delta}_{\rm abs}$, and the maximum absolute error $\Delta_{\rm max}$. Both multiconfigurational SAPT approaches reach similar accuracy. With $\overline{\Delta}_{\rm abs} = 6.0\%$ and $\sigma = 7.4\%$, SAPT(GVB) remains in slightly better agreement with the benchmark than SAPT(CAS) (the respective values for the latter are $\overline{\Delta}_{\rm abs} = 6.9\%$ and $\sigma = 8.9\%$). The error statistics for multiconfigurational SAPT matches SAPT(DFT) results where $\overline{\Delta}_{\rm abs}$ and $\sigma$ amount to $5.6$\% and $7.5$\%, respectively. This indicates a systematic error cancellation between attractive and repulsive energy contributions in the ERPA-based SAPT, since for the individual energy components SAPT(DFT) is clearly closest to SAPT(CCSD).

\begin{table*}
\centering
\caption{Summary of error statistics (in percent) for the SAPT interaction energy for dimers of the TK21/S$_2$ dataset. Errors of the SAPT interaction energy ($E_{\rm int}^{\rm SAPT}$) are given with respect to SAPT(CCSD) results. Errors of the SAPT interaction energies corrected for the $\delta_{\rm HF}$ term ($E_{\rm int}^{{\rm SAPT}+\delta_{\rm HF}}$) are given with respect to CP-corrected~\cite{Boys:70} supermolecular CCSD(T) results calculated in the same basis set. The 2+(CCD) notation refers to the SAPT2+(CCD) scheme. All exchange energy components are computed in the $S^2$ approximation. The basis set is aug-cc-pVTZ.}
\begin{threeparttable}
\begin{tabular}{l S S S S S S S}
\hline\hline
\multicolumn{1}{l}{$\tnote{\textdagger} \;  E_{\rm int}^{\rm SAPT}$} & \multicolumn{1}{c}{HF} & \multicolumn{1}{c}{GVB}  & \multicolumn{1}{c}{CAS}   & \multicolumn{1}{c}{PBE0} & \multicolumn{1}{c}{2+(CCD)} & \\ \cline{2-7}
$\overline{\Delta}$ & -9.87 & -0.80 & -0.44 &  2.72 & -6.95 & \\
$\sigma$            & 14.94 &  7.41 &  8.87 &  7.54 &  5.71 & \\
$\overline{\Delta}_{\rm abs}$ & 12.66 & 5.95 &  6.89 &  5.60 &  7.93 & \\
$\Delta_{\rm max}$  &  33.66 & 12.32 & 14.74 & 16.93 & 14.78 & \\
 &  & &  & &  & \\
\multicolumn{1}{l}{$\tnote{\textdaggerdbl} \; E_{\rm int}^{{\rm SAPT}+\delta_{\rm HF}}$} & \multicolumn{1}{c}{HF} & \multicolumn{1}{c}{GVB}  & \multicolumn{1}{c}{CAS}  & \multicolumn{1}{c}{PBE0} & \multicolumn{1}{c}{2+(CCD)} & \multicolumn{1}{c}{CCSD} \\ \cline{2-7}
$\overline{\Delta}$ & -10.00 & -2.71  & -1.59 &  0.86 & -7.91 & -1.78 \\
$\sigma$            &  10.38 &  6.22  &  7.44 &  7.09 &  5.48 &  3.57 \\
$\overline{\Delta}_{\rm abs}$ & 11.33 & 5.89  &  6.40 &  4.72 &  8.74 & 3.22 \\
$\Delta_{\rm max}$ & 27.64 & 12.08 & 15.69 & 17.24 & 16.46 & 7.96 \\
\hline\hline
\end{tabular}
\begin{tablenotes}
\item[\textdagger] errors with respect to SAPT(CCSD)
\item[\textdaggerdbl] errors with respect to supermolecular CCSD(T)
\end{tablenotes}
\end{threeparttable}
\label{tab:eint}
\end{table*}

It is interesting to compare SAPT interaction energies against the supermolecular CCSD(T) reference. To this end, we approximate higher-order induction effects with the $\delta_{\rm HF}$ term, Eq.~\eqref{eq:delta}. As expected, SAPT(CCSD) is the front runner ($\overline{\Delta}_{\rm abs} = 3.2$\%) followed by SAPT(DFT) with mean absolute error of 4.7\% (lower section of Table~\ref{tab:eint}). Both SAPT(GVB) and SAPT(CAS) are less accurate---the $\overline{\Delta}_{\rm abs}$ value for the former reaches $5.9$\%, while for the latter it amounts to $6.4$\%. Still, the  multiconfigurational SAPT variants outperform not only the Hartree-Fock-based scheme ($\overline{\Delta}_{\rm abs}$ = $11.3$\%), but also the SAPT2 model with the CCD+ST(CCD) dispersion ($\overline{\Delta}_{\rm abs}$ = $8.7$\%). The relatively large errors of SAPT2+(CCD) can be traced to poor representation of the second-order exchange components. Recall that the presented SAPT results neglect exchange effects beyond the $S^2$ approximation which is expected to worsen the agreement between SAPT and CCSD(T) interaction energies.

To summarize, the examined SAPT(GVB) and SAPT(CAS) methods benefit from a partial recovery of the intramonomer correlation effects by the underlying multiconfigurational wave function, as evidenced by a systematic improvement of all energy components with respect to the SAPT(HF) results. Nevertheless, the observed effect is small and relatively large errors compared to fully correlated SAPT schemes persist. This is best exemplified by first-order energies which probe the quality of the monomers density ($E^{(1)}_{\rm elst}$) and density matrices ($E^{(1)}_{\rm exch}$). In the second order the accuracy of  SAPT(MC) is affected both by the missing intramonomer correlation and approximations in the ERPA response equations
(see also discussion in Refs.~\onlinecite{Hapka:19a} and \onlinecite{Hapka:19b}). For the TK21 dataset we observed that both SAPT(GVB) and SAPT(CAS) tend to underestimate second-order contributions which leads to a fortuitous error cancellation in the total interaction energy. When both TK21 and A24 datasets are considered (Table~S24), SAPT(MC) predicts interaction energies with mean absolute errors and standard deviation below $8$\% and $10$\% , respectively, which is significantly better than Hartree-Fock based SAPT ($\overline{\Delta}_{\rm abs} \leq 18$\% and $\sigma \leq 20$\%) and comparable to the SAPT2+(CCD) model ($\overline{\Delta}_{\rm abs} \leq 10$\% and $\sigma \leq 10$\%). 

\section{Conclusions \label{sec:concl}}

We have proposed a SAPT(MC) formalism applicable to dimers in which at least one of the monomers warrants a multireference description. In the approach the interaction energy is expanded through the second-order terms in the intermolecular interaction operator. Formulas for the exchange energy contributions are given in the single-exchange approximation (the $S^2$ approximation) and are valid for ground and nondegenerate excited states of the monomers in spin singlet states. While singlet states require spin-free reduced density matrices, extension to higher spin states is straightforward and involves spin-resolved components of RDMs. Response properties which enter the density-matrix-based SAPT formulas are obtained by solving the extended random phase approximation (ERPA) eigenproblems for each subsystem. Combined with ERPA equations, the presented variant of SAPT requires access only to one- and two-electron reduced density matrices of the monomers. Note that, contrary to the supermolecular method, in SAPT the dimer wave function is never computed which is advantageous for multiconfigurational systems. In this work we applied  SAPT(MC) either with CASSCF or GVB wave functions.

Based on the model \ce{H2}$\cdots$\ce{H2} dimer in which one of the monomers undergoes dissociation, we have verified that SAPT(MC) is capable of describing interactions in systems dominated by nondynamic correlation.
The interaction energy curve from SAPT(GVB) calculations has the correct shape and the largest deviation from the FCI benchmark does not exceed $13$\%. In the \ce{H2}$\cdots$\ce{H2} dimer several active orbitals are sufficient to recover both intra- and intermonomer correlation effects. SAPT(CAS), with only 5 active orbitals per monomer, predicts total interaction energy as well as individual energy contributions with errors below $3$\% with respect to the FCI results. In contrast, SAPT schemes based on single-reference description of the monomers, SAPT(HF) and SAPT(DFT), fail dramatically when entering the strongly-correlated regime.

The proposed multiconfigurational SAPT method is the only one among the existing SAPT approaches that offers the analysis of noncovalent interactions in systems involving electronically excited molecules in singlet states. In this work we examined the role of negative transitions in the linear response function of an excited subsystem in the description of the second order components of SAPT. In Section~\ref{sec:ExcSt} a general protocol for direct evaluation of negative-transition terms has been proposed and its implementation in the ERPA-approximation framework has been presented. As an example of an excited-state complex, we have selected the \ce{C2H4}$\cdots$\ce{Ar} dimer and described it with a small CAS wave function. While for the ground state of the system SAPT(CAS) remains in excellent agreement with supermolecular CCSD(T) results, the excited $\pi \rightarrow \pi^*$ state of ethylene poses a significant challenge. First, we have demonstrated that second-order energy contributions related to negative excitation energies are sizeable and must be accounted for in SAPT. Second, even for the low-lying valence state of ethylene the lack of higher-order induction terms and restriction to the $S^2$ approximation significantly limit the accuracy of SAPT(CAS) results. To illustrate this, we have presented interaction energy curves obtained in a hybrid approach which recovers induction terms up to infinite order in $\hat{V}$. Indeed, a combination of supermolecular CASSCF and second-order dispersion energy from SAPT(CAS) calculations, which we refer to as the CAS+DISP approach,~\cite{Hapka:20m} outperforms SAPT for the $\pi \rightarrow \pi^*$ state, and remains in excellent agreement with the coupled-cluster reference.

The CAS+DISP hybrid can be viewed as SAPT(CAS) supplemented with a CASSCF analogue of the $\delta_{\rm HF}$ term, i.e., the $\delta_{\rm CAS}$ correction. These two methods become equivalent if the $\delta_{\rm CAS}$ term is computed from formula similar to Eq.~\eqref{eq:delta} with CAS supermolecular energy and SAPT(CAS) energy components. While there is no advantage of SAPT(CAS)+$\delta_{\rm CAS}$ procedure over CAS+DISP when both employ the same CAS wave functions, using CAS functions of different levels could be beneficial. Such an approach would employ CAS in the minimal active space to evaluate the $\delta_{\rm CAS}$ term and higher level CAS for description of monomers in SAPT(CAS). Similar to $\delta_{\rm HF}$ correction,~\cite{Patkowski:06,Patkowski:10,Lao:15} addition of $\delta_{\rm CAS}$ would be recommended not only for excited-state complexes but also for ground state polar systems.

To better characterize the performance of SAPT(MC) for many-electron systems, we compared different SAPT schemes against a standard single-reference data set of noncovalently bound dimers. The individual energy components from both SAPT(GVB) and SAPT(CAS) calculations are more accurate than their SAPT(HF) counterparts. This holds also for total interaction energies, where we observe a partial error cancellation between polarization and exchange terms in the second order. The correlated SAPT schemes included in the comparison, i.e., SAPT(DFT) and SAPT2+(CCD), are systematically better than our multiconfigurational SAPT which should be attributed to two factors. One is that ERPA-based SAPT misses the majority of dynamical correlation within the monomers as a result of employing GVB or CAS wave functions with small active spaces. Second is the quality of response properties (exitation energies and transition density matrices) from ERPA equations. Unlike the full linear response (LR-MCSCF,~\cite{Olsen:85} equivalent to MCRPA\cite{Oddershede:84}), ERPA includes response of the orbitals only.

The proposed formulation of multireference SAPT can be applied with wave function methods capable of handling large active spaces, such as density-matrix renormalization group (DMRG)\cite{Legeza:08,Chan:11}, generalized active space (GAS)\cite{Olsen:88,Gagliardi:11} or v2RDM-driven CAS.~\cite{DePrince:16} 
An efficient alternative is offered by range-separated multiconfigurational DFT.~\cite{Savin:95,Fromager:13} 

Without additional approximations SAPT(MC) scales with the sixth power of the molecular size. The computational bottlenecks are the solution of the full ERPA eigenproblem and evaluation of the exchange-dispersion energy formula,~\cite{Hapka:19b} both involving steps with a $n_{\rm OCC}^3 n_{\rm SEC}^3$ cost, where $n_{\rm OCC} = M_{s_1} + M_{s_2}$ and $n_{\rm SEC} = M_{s_2} + M_{s_3}$.

A feasible path to reduce the scaling and increase the efficiency of the method with no damage to the accuracy involves density fitting or Cholesky decomposition techniques routinely applied in single-reference SAPT approaches.~\cite{Misquitta:03,Hesselmann:05,Bukowski:05,Podeszwa:06,Hohenstein:10,Hohenstein:10b,Garcia:20}


\begin{acknowledgement}
The authors would like to thank Grzegorz Cha{\l}asi{\'n}ski for helpful discussions and commenting on the manuscript.

K. P. and M. H. were supported by the National Science Centre of Poland under Grant No. 2016/23/B/ST4/02848.
\end{acknowledgement}

\providecommand{\latin}[1]{#1}
\makeatletter
\providecommand{\doi}
  {\begingroup\let\do\@makeother\dospecials
  \catcode`\{=1 \catcode`\}=2 \doi@aux}
\providecommand{\doi@aux}[1]{\endgroup\texttt{#1}}
\makeatother
\providecommand*\mcitethebibliography{\thebibliography}
\csname @ifundefined\endcsname{endmcitethebibliography}
  {\let\endmcitethebibliography\endthebibliography}{}

\end{document}